%% file: 00_main.tex
\documentclass[amsmath,aip,jcp,floatfix,superscriptaddress,reprint]{revtex4-2}

\usepackage{amssymb}
\usepackage{bm}
\usepackage{breqn}
\usepackage{color}
\usepackage[T1]{fontenc}
\usepackage{graphicx}
\usepackage{hyperref}
\usepackage[utf8]{inputenc}
\usepackage{mathtools}
\usepackage{multirow}
\usepackage{mwe}
\usepackage{placeins}
\usepackage{tabularray}
\usepackage{xr}

\newcommand{\angstrom}{\textup{\AA}}

\newcommand{\citeN}{}
\DeclareRobustCommand*{\citeN}[1]{%
    \begingroup
        \romannumeral-`\x
        \setcitestyle{numbers}%
        \cite{#1}%
    \endgroup
}

\DeclareMathOperator{\e}{e}

\makeatletter
\let\cat@comma@active\@empty
\makeatother

\begin{document}

\title{Dielectric permittivity of water confined in stacks of charged lipid layers: extracting profiles from molecular dynamics simulations using a modified Poisson-Boltzmann equation}

\author{Ludovic Gardré}
\affiliation{Université Claude Bernard Lyon 1, CNRS, Institut Lumière Matière, UMR5306, Villeurbanne, France}

\author{Swen Helstroffer}
\affiliation{UPR 22/CNRS, Institut Charles Sadron, Université de Strasbourg, 23 rue du Loess, BP 84047 67034 Strasbourg Cedex 2, France}

\author{Pierre Muller}
\affiliation{UPR 22/CNRS, Institut Charles Sadron, Université de Strasbourg, 23 rue du Loess, BP 84047 67034 Strasbourg Cedex 2, France}

\author{Fabrice Thalmann}
\affiliation{UPR 22/CNRS, Institut Charles Sadron, Université de Strasbourg, 23 rue du Loess, BP 84047 67034 Strasbourg Cedex 2, France}

\author{Thierry Charitat}
\affiliation{UPR 22/CNRS, Institut Charles Sadron, Université de Strasbourg, 23 rue du Loess, BP 84047 67034 Strasbourg Cedex 2, France}

\author{Laurent Joly}
\affiliation{Université Claude Bernard Lyon 1, CNRS, Institut Lumière Matière, UMR5306, Villeurbanne, France}

\author{Claire Loison}
\email[Corresponding author: ]{claire.loison@univ-lyon1.fr}
\affiliation{Université Claude Bernard Lyon 1, CNRS, Institut Lumière Matière, UMR5306, Villeurbanne, France}

\begin{abstract}
\emph{\textbf{Abstract —}} Most organic and inorganic surfaces (e.g., glass, nucleic acids or lipid membranes) become charged in aqueous solutions. The resulting ionic distribution induces effective interactions between the charged surfaces. Stacks of like-charged lipid bilayers immersed in multivalent ion solutions exhibit strong coupling (SC) effects, where ion correlations cause counter-intuitive membrane attraction. A similar attraction observed with monovalent ions is explained by SC theory through reduced dielectric permittivity under confinement. To explore this phenomenon, we propose a modified Poisson-Boltzmann (mPB) model with spatially varying dielectric permittivity and explicit Born solvation energy for ions. We use the model to investigate the dielectric permittivity profile of confined water in molecular dynamics simulations of charged lipid layers stacks at varying hydration levels, and compare the results with alternative computational methods. The model captures a sharp decrease in permittivity upon dehydration, converging to a plateau value that we attribute to lipid headgroups. The generic nature of the mPB framework allows application to other systems, such as other biological interfaces or solid walls, provided ions follow Boltzmann statistics. Finally, the increase of the area per lipid in our tension-free simulations of the fluid membranes hints that the permittivity decrease upon dehydration is concomitant with an intermembrane attraction.

\end{abstract} \hspace{10pt}

\maketitle

\input{01_intro}

\input{02_methods}

\input{03_results}

\input{04_conclusion}

\input{05_thanks}

\input{06_references}

\pagebreak
\onecolumngrid
\appendix
\setcounter{figure}{0}
\renewcommand{\thesection}{S-\Roman{section}}
\renewcommand{\theequation}{S\arabic{equation}}
\renewcommand{\thefigure}{S\arabic{figure}}
\renewcommand{\theHfigure}{S\arabic{figure}}
\renewcommand{\thetable}{S\arabic{table}}
\renewcommand{\bibnumfmt}[1]{[S#1]}
\renewcommand{\citenumfont}[1]{S#1}

\input{suppMat}

\end{document}

%% file: 01_intro.tex
\section{INTRODUCTION} \label{sec:introduction}

Water is a crucial component of the human body, constituting over half of its composition~\cite{bib:Forbes1953}. In this complex biological system, water exists in confined spaces such as blood capillaries, within cells, or trapped between lipid membranes where it has been extensively studied~\cite{bib:Binder2007}. Confined water exhibits properties significantly different from those in the bulk~\cite{bib:Levinger2002}. The dielectric permittivity in particular is one of these. This macroscopic property of a continuous medium describes how it becomes polarized under an external electric field. The polarization arises from the complex microscopic polarization mechanisms within the medium and generally exhibits non-local behaviour. In bulk water, its value tends to be high due to the polar nature of the molecule. However, when confined, water does not have the freedom to rearrange its structure, thus resulting in a decrease of its dielectric permittivity~\cite{bib:Fumagalli2018}.

\medskip

Electrostatic interactions often govern biological dynamics in the vicinity of membranes~\cite{bib:BenTal1996} (e.g., the binding of proteins to the membranes). Therefore, studying confined water dielectric permittivity is really important to better understand the functioning of living bodies. Furthermore, biological systems generally contain many ions that play an important role in these electrostatic interactions. Depending on the nature of these ions, their number or the system configuration, there can be strong correlation effects between them.

To determine whether these correlation effects must be taken into account, one can compute the coupling constant $\Xi$ of the ions confined between membranes with a given surface charge density $\sigma_{\mathrm{s}}$ as follows~\cite{bib:Moreira2000}:
\begin{equation} \label{eq:couplingConstant}
    \Xi = 2 \pi q^3 \ell_{\mathrm{B}}^2 \frac{\sigma_{\mathrm{s}}}{e} \, ,
\end{equation}
where $e$ is the elementary charge, $q$ is the valence of the ions and $\ell_{\mathrm{B}}$ is the Bjerrum length (the characteristic length at which the thermal energy equals the electrostatic interaction energy between two monovalent ions), given by~\cite{bib:Herrero2024}:
\begin{equation} \label{eq:BjerrumLength}
    \ell_{\mathrm{B}} = \frac{\beta e^2}{4 \pi \varepsilon_0 \varepsilon_r} \, ,
\end{equation}
where $\varepsilon_0$ is the dielectric permittivity of the vacuum, $\varepsilon_r$ is the relative dielectric permittivity of the solvent and $\beta = (k_{\mathrm{B}} T)^{-1}$ with $k_{\mathrm{B}}$ the Boltzmann constant and $T$ the temperature.

\medskip

Low $\Xi$ values correspond to a weak coupling regime where mean-field theories, like the Poisson-Boltzmann model, are adequate. However, these models tend to fail in a strong coupling (SC) regime, i.e. for high $\Xi$ values. SC regime can lead to counter-intuitive phenomena, such as the attraction between stacks of identically charged lipid layers observed by Komorowski and al. in the presence of multivalent counterions~\cite{bib:Komorowski2018}. Depending on the system configuration, SC theory can predict such attraction for $\Xi$ values as low as 10. In the system of Ref.~\citeN{bib:Komorowski2018}, which consists of charged lipid vesicles with Ca\textsuperscript{2+} counterions, $\Xi$ was estimated to be on the order of 20 due to the large valence of the ions (assuming a water dielectric constant of 80), so that SC theory succeeded at explaining the attraction.

\medskip

However, a similar attraction has been experimentally observed by Mukhina and al., this time in the presence of monovalent counterions~\cite{bib:Mukhina2019}. In that situation, it is only possible to have $\Xi$~>~10 by assuming that the dielectric constant of confined water is strongly reduced. For this reason, in the case of monovalent counterions the attraction between identically charged surfaces is perceived as the signature of a strong decrease in the dielectric constant of the confined water~\cite{bib:Schlaich2019}.
Though it is experimentally challenging~\cite{bib:Fumagalli2018, bib:Gramse2013} to probe the dielectric permittivity in these regions, numerical simulations can give good insights of its value. Using the fluctuation-dissipation theorem, Schlaich and al. have shown that one can compute the parallel and perpendicular components of the dielectric permittivity in inhomogeneous systems with no free charges~\cite{bib:Schlaich2016}. Unfortunately, because of this restriction, this method cannot be applied to study confined water in systems containing ions, such as charged lipid membranes, which are very abundant in biology.

\medskip

In this paper, we present an alternative approach to extract an effective dielectric permittivity profile of water confined between stacks of charged lipid membranes, i.e. in the presence of free charges. Our method uses molecular dynamics (MD) simulations and is based on a mean-field approximation with the use of a modified Poisson-Boltzmann (mPB) equation. These modifications combine an effective local $z$-dependant dielectric permittivity~\cite{bib:Huang2008} with the addition of a solvation energy term in the Boltzmann distribution of the ions~\cite{bib:Poddar2016}.
Indeed, because of the differences between lipid-ions  and water-ions solvation, the ions are expelled from the lipid region. This has to be taken into account in the description of the potential energy of the ions~\cite{bib:Schlaich2019}.

In a first part, we present the systems we studied, the details of the mPB model and the methods we used to calibrate and solve it. We then apply the model to investigate the value of the dielectric permittivity of water confined in stacks of charged lipid layers, in the presence of a monovalent salt. We present its results for a large range of hydration level of the lipid membranes, which allow to extract a dielectric permittivity at the centre of the confined water, at each hydration value. We then compare our approach to alternative ways of estimating the dielectric permittivity in such systems.

%% file: 02_methods.tex
\section{METHODS} \label{sec:methods}

\subsection{Studied systems and molecular dynamics simulations} \label{subsec:studiedSystemsMDSimulations}

\begin{figure*}[tbp]
   \centering
   \includegraphics[width=\textwidth]{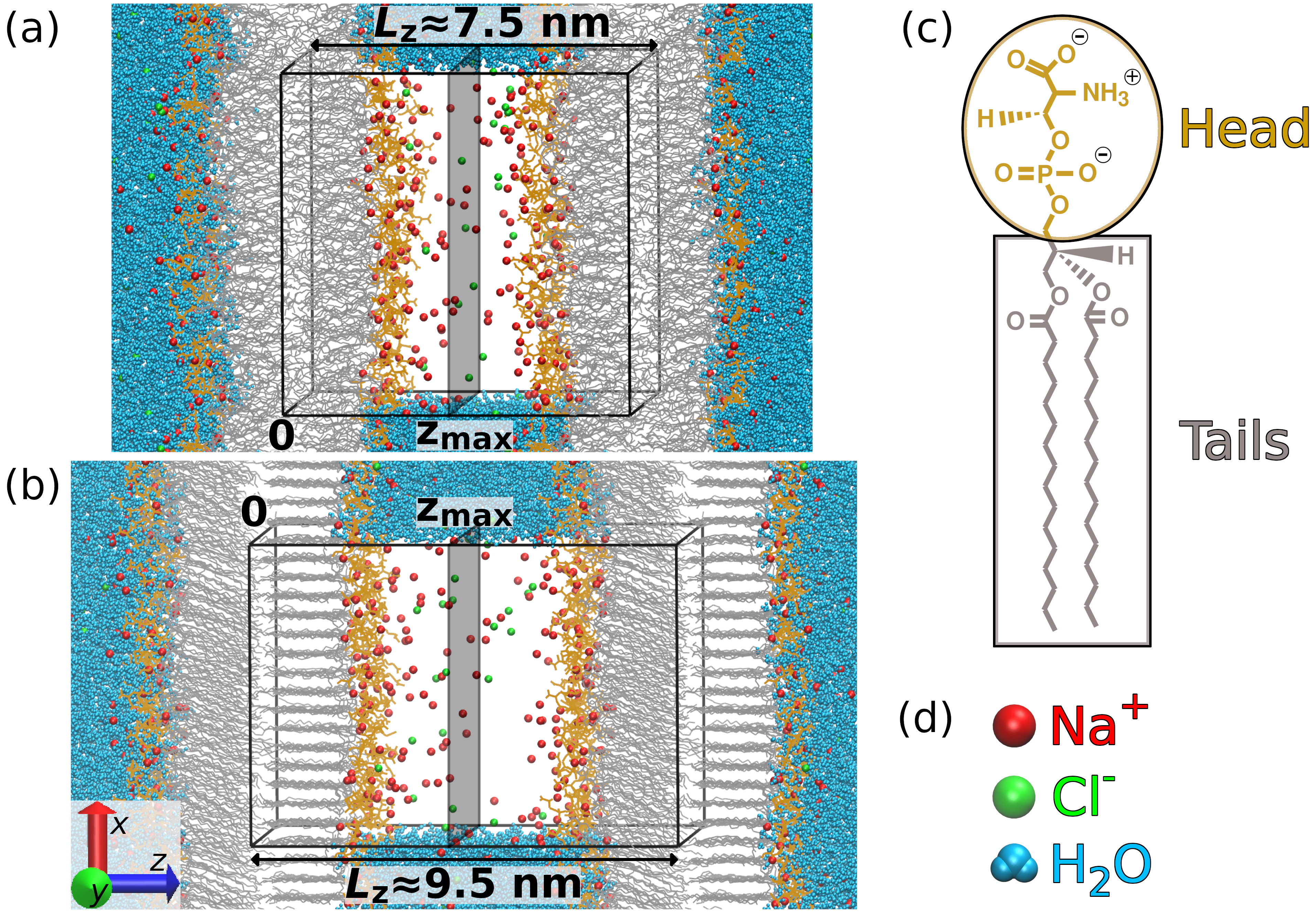}
   \caption{(a, b)~Snapshots of systems at HN~=~35, respectively at $T$~=~353.15\,K (bilayer in the fluid phase) and $T$~=~333.15\,K (bilayer in the gel phase). The blue boxes represent the simulation boxes, replicated in all directions of space. For clarity, the lipids' hydrogen atoms are not drawn in the snapshots, nor are water molecules in the simulation boxes. The $z$ coordinate is set at 0 at the centre of the left bilayer. It reaches $z_{\mathrm{max}}$~=~$L_z / 2$ at the middle of the interlayer water. (c)~Detail of the DPPS lipid molecule. The gold and silver colours respectively correspond to the heads and the tails of the lipids as drawn in the snapshots. (d)~Interlayer species.}
   \label{fig:systemPresentation}
\end{figure*}

We have followed the standard setup for planar bilayers: a tensionless membrane patch with 3-dimensional periodic boundary conditions and the $z$-axis along the bilayer normal. Figs.~\ref{fig:systemPresentation}(a) and (b) show snapshots of example systems at high hydration and at two different temperatures.

\medskip

\textit{Composition} --- 
All systems were built from the same base elements : a lipid bilayer made of 1,2-dipalmitoyl-sn-glycero-3-phosphoserine (DPPS, negatively charged in the head) solvated in water, see Fig.\ref{fig:systemPresentation}.
We neutralized the charge of the lipids with Na\textsuperscript{+} counterions. In addition, to test the model in presence of both cations and anions, we included NaCl salt in the solvent, at a concentration of approximatively 0.25\,M.
The amount of water plays an important role in these systems. It was controlled at the creation step by setting a hydration number (HN), corresponding to the number of water molecules per lipid (e.g., HN~=~20 means there are 20 water molecules per lipid).
For HN~$\geq 11$, the bilayer was made of two leaflets of 100 lipids. At lower HN, to keep the salt concentration constant while still having a statistically relevant number of salt ions in the system, we quadrupled the systems size so each leaflet of the bilayer contained 400 lipids. The compositions of the different systems are detailed in Tab.~\ref{SI:tab:SimulationsList} in suplementary material.

\medskip

\textit{Construction of initial conformations} --- 
For HN~$\geq 11$, the systems were built using charmm-gui~\cite{bib:Jo2008}. This tool provided pre-equilibrated systems with bilayers in the fluid phase. We followed the charmm-gui default pre-equilibration process that consists of six runs: two 125-ps runs in the (N,$V$,$T$) ensemble, followed by one 125-ps run and three 500-ps runs in the (N,$P$,$T$) ensemble. Restraints are applied on the lipid atoms’ positions and on the dihedrals inside the lipid molecules. These restraints are gradually removed run after run throughout the whole equilibration process.
For HN\,$\leq10$,  systems were built iteratively: for each HN, we started by replicating the system at the HN just above (e.g., we replicated the HN~=~11 system to create the HN~=~10 system, and so on down to HN~=~02). We then randomly removed water molecules and/or ions to match the desired HN and salt concentration.

\medskip

\textit{Force field} --- 
We chose CHARMM36~\cite{bib:Klauda2010} for the lipid atoms and  the modified TIP3P (mTIP3P)~\cite{bib:MacKerell1998} for water, since CHARMM36 was originally developed with this water force field. It is known that TIP3P permittivity does not match the experimental value for bulk water, but the typical decrease of permittivity when temperature is increased is reproduced~\cite{bib:Izadi2014}. We performed independent simulations on bulk mTIP3P water, either pure or with 0.25\,M NaCl that confirm this result for the mTIP3P model (see Sect.\,\ref{SI:sec:calibrationDetails} in Supplementary Material). Thus, in this work, we focus on the relative variations of the permittivity, not the absolute values.
We did not use explicit polarisable force fields~\cite{bib:Vanommeslaeghe2015}, nor scaled-charge force fields~\cite{bib:DuboueDijon2020}. These may be considered if one would extend this study~\cite{bib:Melcr2020}.
Moreover, one has to remember that the impact of dehydration on the lipid bilayer structure is not yet perfectly reproduced by CHARMM36~\cite{bib:Kurki2024}.

Short-ranged van der Waals interactions were modelled with a Lennard-Jones potential.
A cutoff distance was set at 12\,\angstrom, and we smoothly switched forces to 0 between 10 and 12\,\angstrom. We used Lorentz-Berthelot combination rules to compute the cross-species parameters.
Long ranged Coulombic interactions were computed with the Particle-Mesh Ewald method~\cite{bib:Essman1995,bib:Darden1993}.
All simulations were carried out with a time step of 2\,fs, using Gromacs 2021.2 with CUDA support~\cite{bib:Abraham2015}. 

\medskip

\textit{Simulation Ensemble} --- 
The equilibration and production runs were performed in the \mbox{(N, $P_n$, $\gamma$, $T$)} ensemble, were $P_n$ is the normal pressure and $\gamma$ is the surface tension.
The temperature was maintained with a Nosé-Hoover thermostat~\cite{bib:Martyna1992}. 
Systems were kept in the \mbox{$P_n=1$\,bar} and $\gamma=0$ ensemble with the semi-isotropic Parrinello-Rahman barostat~\cite{bib:Parrinello1981}, i.e.\,with the constraints \mbox{$L_x = L_y$}, and \mbox{$P_z = (P_x + P_y)/2$}.

\medskip

\textit{Equilibration and Production} --- 
Every system provided by charmm-gui was first run at $T$~=~333.15\,K during 1\,µs. 
Then, we increased the temperature up to $T$~=~353.15\,K with a step of 5\,K. For every temperature, we ran a 1\,µs simulation before increasing the temperature. 

Even if all initial configurations were initially prepared in the fluid phase, depending on the HN and temperature, we observed spontaneous phase transitions from fluid to gel (or vice versa). When it occurred, we extended the simulation length and discarded everything before the transition. Therefore, all observables shown in this work were computed over 1\,µs phase-transition-free production runs. The error bars in Figs.\,\ref{fig:Epsilon_vs_HN}, \ref{fig:EpsilonFromDipoles_vs_HN} and \ref{fig:ApL_vs_HN} were obtained by dividing the production run into five trajectories of 200\,ns, which were analysed separately and considered as independent. The standard error of the mean was multiplied by the appropriate Student's t-factor for a 95\% confidence interval (i.e. $\simeq~2.78$ for five values).

\textit{Lipid ordering} --- 
Depending on its hydration and temperature, the equilibrated bilayer can be in the fluid or  in the gel phase.  At the highest temperature, the bilayer is in the fluid phase and consequently the lipids are disordered, see Fig.\,\ref{fig:systemPresentation}(a). At the lowest temperature, the bilayer is in the gel phase: the lipid tails are ordered, but not their heads, see Fig.\,\ref{fig:systemPresentation}(b).

Due to the structural difference between these two phases, the membrane has a different thickness which influences the box length in the $z$ direction, meaning that $L_z$ is not the same in all systems and varies depending on the physico-chemical conditions (e.g., $T$, HN, etc.).

Even if the orientation of the lipid tilts may be different in the two monolayers in the gel phase, see Fig.\,\ref{fig:systemPresentation}(b), all the systems are considered symmetrical with respect to the plane defined by $z$~=~$z_{\mathrm{max}}$. In practice, when computing charge densities from MD simulations, the centre of mass of the bilayer is placed at the centre of the simulation box (in Fig.~\ref{fig:systemPresentation}, this is equivalent to shifting the simulation box by $L_z/2$ in the $z$ direction). The densities are then integrated along the $z$ axis, then symmetrized and centred with respect to the bilayer centre of mass.

\medskip

\subsection{Modified Poisson-Boltzmann model} \label{subsec:mpbModel}

The purpose of this work is to establish a method to obtain, from MD simulations, an effective profile of the dielectric permittivity of the ion-carrying water confined between charged lipid membranes. To do so, we propose to modify the Poisson-Boltzmann equation, a mean-field approximation that describes how electrolyte solutions behave in the vicinity of a charged interface~\cite{bib:Israelachvili2011}.

\medskip

First, let us recall the standard Poisson-Boltzmann model, starting from Maxwell-Gauss equation in matter:
\begin{equation} \label{eq:MaxwellGauss}
    \bm{\nabla} \cdot \bm{D} = \rho_{\mathrm{free}} \, ,
\end{equation}
where $\bm{D}$ is the electric displacement field and $\rho_{\mathrm{free}}$ is the free-charge density.
Considering a planar geometry (with the $z$ axis perpendicular to the charged plane), the problem is reduced to one dimension as a consequence of the system’s symmetries and invariances:
\begin{equation} \label{eq:Poisson1D}
    -\varepsilon_0 \varepsilon_{\perp} \frac{\mathrm{d^2}V}{\mathrm{d}z^2} = \rho_{\mathrm{free}}(z) = \rho_{\mathrm{ions}}(z) \, ,
\end{equation}
where $V$ is the electric potential, $\varepsilon_{\perp}$ is the perpendicular component of the relative permittivity (assumed homogeneous) given by the relation $\bm{D_{\perp} = \varepsilon_{\perp} \bm{E_{\perp}}}$, and considering the total free-charge density is equal to the charge density of the ions $\rho_{\mathrm{ions}} = \rho_{+} + \rho_{-}$, where $\rho_{+}$ and $\rho_{-}$ are the charge densities of the cations and anions, respectively. Using a Boltzmann distribution of cations and anions:
\begin{equation} \label{eq:rhoPlusMinus}
    \rho_{\pm}(z) = \pm \, e C_{\mathrm{salt}} \exp \left[ \mp \beta e V(z) \right] \, ,
\end{equation}
with $C_{\mathrm{salt}}$ the ions concentration in a reservoir where $V$ vanishes, one can express $\rho_{\mathrm{ions}}$ as follows:
\begin{equation} \label{eq:ionsBoltzmannDistribution}
    \begin{split}
        \rho_{\mathrm{ions}}(z) & = \rho_{+}(z) + \rho_{-}(z) \\
                       & = e C_{\mathrm{salt}} \Bigl\{ \exp \Bigl[ -\beta e V(z) \Bigr] \Bigr. \\
                       & \hphantom{{}= e C_{\mathrm{salt}} \{ } \Bigl. - \exp \Bigl[ \beta e V(z) \Bigr] \Bigr\} \, .
    \end{split}
\end{equation}

Combining Eq.~\eqref{eq:Poisson1D} and Eq.~\eqref{eq:ionsBoltzmannDistribution}, one derives the standard 1D Poisson-Boltzmann equation:
\begin{equation} \label{eq:PoissonBoltzmann1D}
    \frac{\mathrm{d^2}V}{\mathrm{d}z^2} = \frac{2 e C_{\mathrm{salt}}}{\varepsilon_0 \varepsilon_{\perp}} \sinh \left[ \beta e V(z) \right] \, .
\end{equation}

Introducing the dimensionless reduced electric potential $\phi$, defined as $\phi(z) = \beta e V(z)$, and the Debye length $\lambda_{\mathrm{D}}$ (the characteristic length at which the electrolyte solution screens the electric field):
\begin{equation} \label{eq:debyeLength}
    \lambda_{\mathrm{D}} = \sqrt{\frac{\varepsilon_0 \varepsilon_{\perp}}{2 \beta e^2 C_{\mathrm{salt}}}} \, ,
\end{equation}
one can rewrite Eq.~\eqref{eq:PoissonBoltzmann1D} as:
\begin{equation} \label{eq:PoissonBoltzmann1D_withReducedPotential}
    \frac{\mathrm{d^2}\phi}{\mathrm{d}z^2} = \frac{1}{\lambda_{\mathrm{D}}^2} \sinh \left[ \phi(z) \right] \, .
\end{equation}

\medskip

In our modified Poisson-Boltzmann (mPB) model, we allowed the relative permittivity $\varepsilon_{\perp}$ to vary along the $z$ axis~\cite{bib:Huang2008}. In addition, we included the lipid heads charge density $\rho_{\mathrm{lipids}}$ as a fixed external contribution to the total free-charge density. We then rewrote Eq.~\eqref{eq:Poisson1D} as follows:
\begin{equation} \label{eq:modifiedPoisson1D}
    \begin{split}
        -\varepsilon_0 \frac{\mathrm{d}}{\mathrm{d}z} \left[ \varepsilon_{\perp}(z) \frac{\mathrm{d}V}{\mathrm{d}z} \right] & = \rho_{\mathrm{free}}(z)\\
                                                   & = \rho_{\mathrm{lipids}}(z) + \rho_{\mathrm{ions}}(z) .
    \end{split}
\end{equation}

\medskip

Since $\varepsilon_{\perp}$ varies with $z$, so does the ions solvation energy. We then introduced the solvation energy  $W_{\pm}$ in the Boltzmann factor, using the Born expression~\cite{bib:Poddar2016} :
\begin{equation} \label{eq:BornEnergy}
    W_{\pm}(z) = \frac{e^2}{8 \pi \varepsilon_0 r_{\pm}} \left( \frac{1}{\varepsilon_{\perp}(z)} - \frac{1}{\varepsilon_{\perp}^{\mathrm{ref}}} \right) \, ,
\end{equation}
where $r_{\pm}$ is the radius of the ion and $\varepsilon_{\perp}^{\mathrm{ref}}$ is a reference value for the relative permittivity of water (e.g., bulk water value).
This potential describes the energetic cost to create an electric field in the surrounding dielectric medium due to the presence of the ion. It represents the energy for the dipoles in the solvent to reorient because of the presence of the charge. The increase of this energy when the ion enters the lipid region describes that the lipid heads and tails are more difficult to polarize than the water dipoles. This Born term therefore represents how the the lipid-ion solvation differs from the water-ion solvation. Noticeably, it approximates solely dipole/ion electrostatic interactions.  The cost to create a cavity for the ion is not included in this Born term. This model therefore considers that the free energy associated to the Lennard-Jones interactions of the ions remain approximatively constant when the ions move relative to the lipidic membrane.
The values of the parameters  of the Born model will be discussed in section \ref{subsec:calibratinMpbModel}, devoted to the model calibration. In the end, the ions charge density became:
\begin{equation} \label{eq:modifiedIonsBoltzmannDistribution}
    \begin{split}
        \rho_{\mathrm{ions}}(z) & = \rho_{+}(z) + \rho_{-}(z) \\
                       & = e C_{\mathrm{salt}} \\
                       & \times \Bigl\{ \exp \Bigl[ -\beta W_{+}(z) - \beta e V(z) \Bigr] \Bigr. \\
                       & \hphantom{{}= \, } \Bigl. - \exp \Bigl[ -\beta W_{-}(z) + \beta e V(z) \Bigr] \Bigr\} \, .
    \end{split}
\end{equation}

\subsection{Solving the mPB model's equation} \label{subsec:solvingMpbModel}

In this section, we introduce the method we followed to solve the mPB model shown in Eq.~\eqref{eq:modifiedPoisson1D}. In this equation, the ions charge density $\rho_{\mathrm{ions}}$ is given by the analytical expression from Eq.~\eqref{eq:modifiedIonsBoltzmannDistribution}, while the lipid charge density $\rho_{\mathrm{lipids}}$ corresponds to values extracted from MD simulations beforehand. These values were centred and symmetrized around the centre of mass of the bilayer, allowing to focus on one single half of the system when solving the equation. Therefore, we set the origin of the $z$ axis at the centre of the bilayer and the maximum $z$ value then corresponded to the centre of the interlayer water channel.

\medskip

There were two unknown profiles in the equation: the electric potential $V(z)$ and the relative permittivity $\varepsilon_{\perp}(z)$.
We chose to model the latter with a parametrized curve. This led to having only one fully unknown profile, the electric potential, so we could numerically solve the equation.

\medskip

To decide on the shape of the dielectric permittivity profile, we analysed a system made of a zwitterionic lipid bilayer and no ions, i.e. without any free charges. In this situation, the electric displacement field is constant in the solvent, and the dielectric response can be obtained as the linear response of the polarization density upon the perturbation of an
electric field. Fluctuation-dissipation approaches can then be used to extract a dielectric permittivity profile from the local dipole moment fluctuations in equilibrium simulations~\cite{bib:Schlaich2019}. We used MAICoS (\url{https://www.maicos-analysis.org}), which provided us a profile of $\varepsilon_{\perp}^{-1}(z)$ that was close to a sigmoid (see Fig.~\ref{SI:fig:epsilon_perp}). This shape was similar to the inverse permittivity profiles obtained by Schlaich et al. for neutral lipid bilayers such as digalactosyldiacylglycerol (DGDG)~\cite{bib:Loche2019}.
Supposing that the shape of the dielectric permittivity profile would be similar in a lipidic system with free charges, we chose to model its inverse (to stay consistent with the reference provided by MAICoS) using one sigmoid and one Gaussian, to allow for more degrees of freedom:
\begin{equation} \label{eq:epsilonPerpInvModel}
    \begin{split}
        & \varepsilon_{\perp}^{-1} \left( z, z_1, K, \sigma, \mu, C, \varepsilon_{\perp}^{\mathrm{tails}}, \varepsilon_{\perp}^{\mathrm{plateau}} \right) \\
        = & \, \frac{1}{\varepsilon_{\perp}^{\mathrm{plateau}}} + \left( \frac{1}{\varepsilon_{\perp}^{\mathrm{tails}}} - \frac{1}{\varepsilon_{\perp}^{\mathrm{plateau}}} \right) \\
        & \hphantom{{{}}= \frac{1}{\varepsilon_r^{\mathrm{plateau}}} } \times \frac{1}{1 + \exp \Bigl[ K \left( z - z_1 \right) \Bigr]}  \\
        & + \frac{C}{\sigma \sqrt{2 \pi}} \exp \left[ - \frac{\left( z - \mu \right)^2}{2 \sigma^2} \right] \, .
    \end{split}
\end{equation}
$z_1$ and $K$ are parameters for the sigmoid component: the first one controls the coordinate of its inflection point while the other one controls its stiffness. $\sigma$, $\mu$ and $C$ are parameters for the Gaussian component. The first two respectively correspond to its standard deviation and its expected value, while $C$ controls its height. Finally, $\varepsilon_{\perp}^{\mathrm{tails}}$ and $\varepsilon_{\perp}^{\mathrm{plateau}}$ respectively control the left and the right horizontal asymptotes. An example of this curve is shown in Fig.~\ref{SI:fig:mPB_333K} in the supplementary material. The effect of the addition of a second Gaussian function in Eq.~\ref{eq:epsilonPerpInvModel} is also displayed in Fig.~\ref{SI:fig:mPB_353K_gaussians}. We considered the results with a single Gaussian more robust.

Equation~\ref{eq:epsilonPerpInvModel} reflects our choice for a profile without large or numerous oscillations. It cannot describe complex profiles with several oscillations or negative values, as the ones that have been calculated for flat interfaces such as graphene~\cite{bib:Loche2019}. 
Indeed, we expect that the roughness of the lipidic interface  diminishes strong density oscillations and associated permittivity oscillations. 
Moreover, our definition for $\varepsilon_{\perp}^{-1}$ is thought as a "coarse-grained dielectric constant", as the one proposed by Borgis et al.~\cite{bib:Borgis2023}. This quantity (noted $\tilde{\varepsilon}_{\perp}(z)$ in their article) is averaged over slices of the thickness of the same size as a water molecule. Borgis et al. show that $\tilde{\varepsilon}_{\perp}$ can be defined locally, and that the $z$-averaging leads to smooth profiles, without strong oscillations.

\medskip

Since we parametrized the inverse of the dielectric permittivity, we modified Eq.~\eqref{eq:modifiedPoisson1D} to obtain an equation depending on $\varepsilon_{\perp}^{-1}$ and its derivative with respect to $z$. The complete system of equations numerically solved is  given in section \ref{SI:sec:numericalSolving} in the supplementary material.

\medskip

In this solving process, we added an optimization procedure, to obtain the best values for the parameters of $\varepsilon_{\perp}^{-1}$ (Eq.~\eqref{eq:epsilonPerpInvModel}). To do so, we fitted the ionic charge densities computed with Eq.~\eqref{eq:modifiedIonsBoltzmannDistribution} on the ionic charge densities extracted from the MD simulation.

\subsection{Calibrating the mPB model} \label{subsec:calibratinMpbModel}

The calibration process served the purpose of fixing the value of two parameters: $C_{\mathrm{salt}}$ and $\varepsilon_{\perp}^{\mathrm{ref}}$.

\medskip

Three parameters were set prior to the calibration: the ionic radii of Na\textsuperscript{+} and Cl\textsuperscript{-} and the asymptotic value for the relative permittivity in the lipid tails $\varepsilon_{\perp}^{\mathrm{tails}}$~=~2. This later choice should be discussed. Values obtained from MD simulations are usually reported around 1~\cite{bib:Stern2003,bib:Nymeyer2008}, in agreement with the value we obtain in the analysis of the DPPC bilayer with MAICoS, as visible in Fig.~\ref{SI:fig:epsilon_perp}. However, theoretical  calculation~\cite{bib:Huang1977} and experimental measurements~\cite{bib:Gramse2013} report a value closer to 2. Fig.~\ref{SI:fig:epsilonR_tails} shows that this choice has practically no impact on the permittivity profile in the region where ions are present.

The Born radii are in fact effective parameters, and may differ from the commonly measured ionic crystal radius~\cite{bib:Rashin1985,bib:Kournopoulos2022,bib:Silva2024}. We have used the Born radii of these ions calculated from their free energies of hydration by Schmid and al.~\cite{bib:Schmid1999}: $r_{+}$~=~0.187\,nm and $r_{-}$~=~0.186\,nm, after checking that the values chosen had no significant impact on the results.

\medskip

We calibrated the model for each of the five temperatures. Every calibration was realised in two stages, using the HN~=~45 system which contained enough water to have a solvent reservoir between the membranes. The first stage was to determine the ions concentration in the reservoir $C_{\mathrm{salt}}$. Using the ionic Boltzmann distributions as in Eq.~\eqref{eq:modifiedIonsBoltzmannDistribution}, one can get rid of the electrostatic energy by multiplying the cations and anions concentrations (respectively $C_{+}$ and $C_{-}$):
\begin{equation} \label{eq:ionsConcentrationProduct}
    \begin{split}
        C_{+}(z) \times C_{-}(z) & = C_{\mathrm{salt}} \exp \Bigl[ - \beta e V(z) - \beta W_{+}(z) \Bigr] \\
        & \hphantom{{}= \, } \times C_{\mathrm{salt}} \exp \Bigl[ \beta e V(z) - \beta W_{-}(z) \Bigr] \\
        & = C_{\mathrm{salt}}^2 \exp \Bigl[ - \beta \Bigl( W_{+}(z) + W_{-}(z) \Bigr) \Bigr] \, .
    \end{split}
\end{equation}

In the reservoir, the ions concentration $C_{\mathrm{salt}}$ is constant. Moreover, one can also set $W_{\pm}$~=~0 in the reservoir by defining the relative permittivity of the reservoir water as the reference value $\varepsilon_{\perp}^{\mathrm{ref}}$ (Eq.~\eqref{eq:BornEnergy}), which will be obtained in the second stage of the calibration. In the reservoir, Eq.~\eqref{eq:ionsConcentrationProduct} then gives:
\begin{equation} \label{eq:CsaltCalibration}
    C_{\mathrm{salt}} = \sqrt{C_{+} \times C_{-}} \, .
\end{equation}

Fig.~\ref{SI:fig:mPB_calibration} in the supplementary material shows the ionic concentrations $C_{+}(z)$  and $C_{-}(z)$ in the most hydrated system at $T$~=~353.15\,K, and the square root of their product that indeed reaches a constant value in the reservoir. The value of $C_{\mathrm{salt}}$ was then obtained using Eq.~\ref{eq:CsaltCalibration} in the reservoir, i.e. at large $z$ values. The values  $C_{\mathrm{salt}}$ are reported Tab.~\ref{SI:tab:calibrationValues}. They are all close to $0.20\pm0.01$~M.
\medskip

Once this first step was done, we solved the mPB model's equation by following the method described in the precedent section \ref{subsec:solvingMpbModel}, except we added the following constraint during the optimization process: $\varepsilon_{\perp}^{\mathrm{plateau}}~=~\varepsilon_{\perp}^{\mathrm{ref}}$. The dielectric permittivity indeed reached a plateau in the water region (where $z$ values are close to $z_{\mathrm{max}}$, as shown in Fig.~\ref{fig:systemPresentation}). We identified this plateau value, which corresponds to the horizontal asymptote defined by the parameter $\varepsilon_{\perp}^{\mathrm{plateau}}$ in Eq.~\eqref{eq:epsilonPerpInvModel}, as the dielectric permittivity of the reservoir water. Since this parameter is optimized, the constraint allowed to maintain $W_{\pm}$~=~0 in the reservoir, ensuring self-consistency with the first stage of the calibration procedure.

Finally, we kept the optimized value of $\varepsilon_{\perp}^{\mathrm{plateau}}$ as the value of $\varepsilon_{\perp}^{\mathrm{ref}}$ in all the systems at the same temperature. The values  $\varepsilon_{\perp}^{\mathrm{ref}}(T)$ for all temperatures are reported in Supplementary Material, Tab.~\ref{SI:tab:calibrationValues}.
One expects these values of $\varepsilon_{\perp}^{\mathrm{ref}}(T)$ to approach the values of bulk mTIP3 at the relevant temperature and $C_{\mathrm{salt}}$ salt concentration.  The permittivity of bulk TIP3P water is known to decrease with increasing temperature~\cite{bib:Izadi2014}. We observe the same tendency for our values of 
$\varepsilon_{\perp}^{\mathrm{ref}}(T)$.
The permittivity of bulk TIP3P water is also known to decrease  with the addition of salt~\cite{bib:Zhang2023}.  We also observed the same tendency for our values of $\varepsilon_{\perp}^{\mathrm{ref}}$, that are about 10 to 15\% lower than the values for pure mTIP3P at the given temperature, measured in independent simulations (see \mbox{Sect.~\ref{SI:sec:calibrationDetails}}). We conclude that our reference permittivities for the reservoirs are reasonable.

%% file: 03_results.tex
\section{RESULTS}

\subsection{Solutions of the mPB model and corresponding profiles}

\begin{figure*}[tbp]
   \centering
   \includegraphics[width=\textwidth]{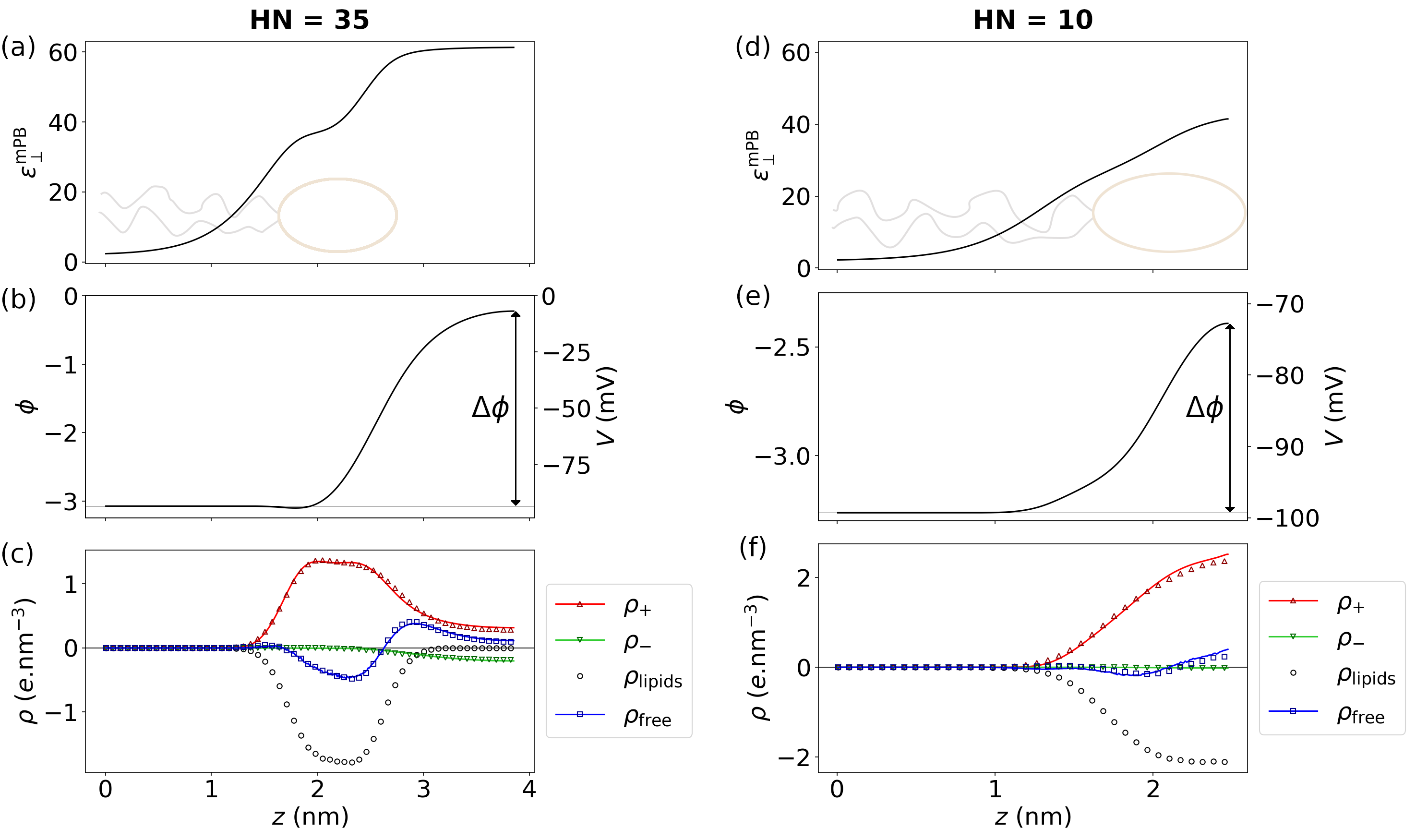}
   \caption{Results from the modified Poisson Boltzmann model at 353.15\,K, at HN~=~35 (left) and HN~=~10 (right). (a, d)~Relative dielectric permittivity profile. The lipid in the background represents the approximative $z$ position of the tails and the heads in the lipid layer. (b, e) Reduced electric potential $\phi(z)$ and electric potential $V(z)$. (c, f) Charge densities. Markers show charge densities extracted from the MD simulation (for clarity, only one data point out of seven is plotted). Lines show charge densities obtained with the mPB model, i.e. with Eq.~\eqref{eq:modifiedIonsBoltzmannDistribution}. According to Eq.~\eqref{eq:modifiedPoisson1D}, both blue plots include the lipid charge density extracted from the MD simulation.}
   \label{fig:mPB_353K}
\end{figure*} 

After the calibration was done, we were able to solve the mPB model's equation (Eq.~\eqref{SI:eq:mPB_ODE_system_with_inverse_permittivity} in the supplementary material). For each system, we obtained three profiles: the electric potential $V(z)$, the dielectric permittivity $\varepsilon_{\perp}^{\mathrm{mPB}}(z)$ and the ionic charge densities $\rho_{\pm}(z)$ predicted by the model (computed using the two latter and Eq.~\eqref{eq:modifiedIonsBoltzmannDistribution}). Examples of these results for a DPPS bilayer at 353.15\,K are shown in Fig.~\ref{fig:mPB_353K}, both at high hydration (HN~=~35, left side) and low hydration (HN~=~10, right side). As illustrated in Fig.~\ref{fig:systemPresentation}, thanks to the symmetry of the system, all plots are simplified to show only one half of the system. The lipid tails are located in the left side, the lipid heads are in the region where $\rho_{\mathrm{lipids}}$ is not null and the water is on the right side (with an overlap in the hydrophilic heads).

\medskip

Fig.~\ref{fig:mPB_353K}\,(a,\,d) show the estimated relative dielectric permittivity profiles $\varepsilon_{\perp}^{\mathrm{mPB}}$ in the systems, obtained by inverting Eq.~\eqref{eq:epsilonPerpInvModel}. The general shape of the profiles naturally depends on the parametrized curve we chose to describe its inverse: one sigmoid and one Gaussian. In particular, thanks to the Gaussian component, the dielectric permittivity can display a small bump. This can notably be the case in the lipid heads region, where the structure is more complex because of the presence of free charges (ions) and the complex distribution of charges and dipoles in the lipid heads. An example solution of the mPB model without the Gaussian component is given in Fig.~\ref{SI:fig:mPB_333K} in the supplementary material to demonstrate its importance.

At high hydration number, the dielectric permittivity reaches a plateau in the confined interlayer water. This confirms that the bilayer is highly hydrated, therefore at the centre of the channel the water molecules are not constrained by the interactions with the lipids. With further dehydration however, this plateau is never reached, as seen in Fig.~\ref{fig:mPB_353K}\,(d). In this hydration state, all the remaining water molecules are located in the lipid heads region, where their orientation is somewhat constrained by electrostatic interactions, which is why a plateau is never reached.

\medskip

Since the system is considered symmetrical with respect to the plane defined by $z = 0$ (other half of the bilayer, as visible in Fig.~\ref{fig:systemPresentation}) and periodic through the MD periodic boundary conditions, we imposed when solving the model's equation that the derivative of the reduced electric potential is null at the borders to ensure its continuity, as visible in Fig.~\ref{fig:mPB_353K}\,(b, e).

For highly hydrated systems, we found a magnitude of about 100\,mV for the electric potential difference $\Delta V$ between the lipid heads and the water, which corresponds to $\Delta \phi$~$\approx$~3.3. Though we did not manage to find an other value for a DPPS membrane in the literature to compare with the value given by the model, we found some values for a 1,2-dipalmitoyl-sn-glycero-3-phosphocholine (DPPC) membrane. This lipid has a different head than that of DPPS and is zwitterionic instead of being negatively charged. Stern and Feller carried out simulations of DPPC membranes~\cite{bib:Stern2003} and found a magnitude of around 1\,V ($\Delta \phi \approx 33$), while experimental values~\cite{bib:Gawrisch1992} gave a smaller magnitude of 227\,mV ($\Delta \phi \approx 7.5$). One of the reasons the authors gave to explain such difference was the inaccuracy of the force field they used, which was CHARMM27~\cite{bib:Feller2000}. Despite being a different lipid, we expected the order of magnitude of DPPS membrane's electric potential difference to be somewhat close from that of DPPC membrane's. To strengthen this hypothesis, one can use Grahame equation~\cite{bib:Israelachvili2011}, which derives from the standard Poisson-Boltzmann model and links the surface charge to the surface reduced potential difference, in the case of a planar solid wall:
\begin{equation} \label{eq:GrahameEquation}
    \sinh \left( \frac{\Delta \phi}{2} \right) = \frac{2 \pi \ell_B \lambda_{\mathrm{D}}}{q} \frac{\sigma_{\mathrm{s}}}{e} \, .
\end{equation}

Applied in our systems, where $\sigma_{\mathrm{s}}$ varies between 1 $e$ per 0.4 to 0.6\,nm\textsuperscript{2} (depending on the area per lipid shown in Fig.~\ref{fig:ApL_vs_HN}), it gives a value $\Delta \phi$~$\approx$~4, which is consistent with the value we measured. Since we used a much more recent version of the CHARMM force field than Stern and al., our results tend to validate their statement against the accuracy of CHARMM27.

\medskip

With the profiles obtained in Fig.~\ref{fig:mPB_353K}\,(a) and (b), the model computes the ionic charge densities using Eq.~\eqref{eq:modifiedIonsBoltzmannDistribution}. The results are shown in Fig.~\ref{fig:mPB_353K}\,(c) and compared with the charge densities obtained from the MD simulations. For both anions and cations, the charge densities computed by the model are in good agreement with MD data. In addition, since the solution for the electric potential appears to be consistent within the PB model, these results suggest that the profile of the dielectric permittivity $\varepsilon_{\perp}^{\mathrm{mPB}}$ is a good approximation of the permittivity in the system.

In comparison to the classical PB model, the ions are more repelled by the interface. In our mPB, this repulsion is due to the solvation energy, via the Born energy term.
Fig.~\ref{SI:fig:energyComparison} shows that the Born energy dominates over the electrostatic potential when entering the lipid region.
In other mPB formulations by Schlaich et al., this repulsion was also taken into account by other forms of potentials~\cite{bib:Schlaich2019}, but they also dominate over the electrostatic energy nearby the interface.

\medskip

Similarly, with the profiles obtained in Fig.~\ref{fig:mPB_353K}\,(d) and (e), the model computes the ionic charge densities shown in Fig.~\ref{fig:mPB_353K}\,(f). This result shows that even in a low hydration state, the charge densities computed by the model stay in good agreement with MD data. Moreover, the model is also able to reproduce more complex shapes of charge densities as they appear in systems where the bilayer is in the gel phase. Fig.~\ref{SI:fig:mPB_333K}\,(a) in the supplementary material shows results from the mPB model on a system at HN~=~35 and $T$~=~333.15\,K, which corresponds to a bilayer in the gel phase.

\subsection{Dielectric permittivity at the centre of the interlayer water channel}

\begin{figure}[tbp]
   \centering
   \includegraphics[width=\columnwidth]{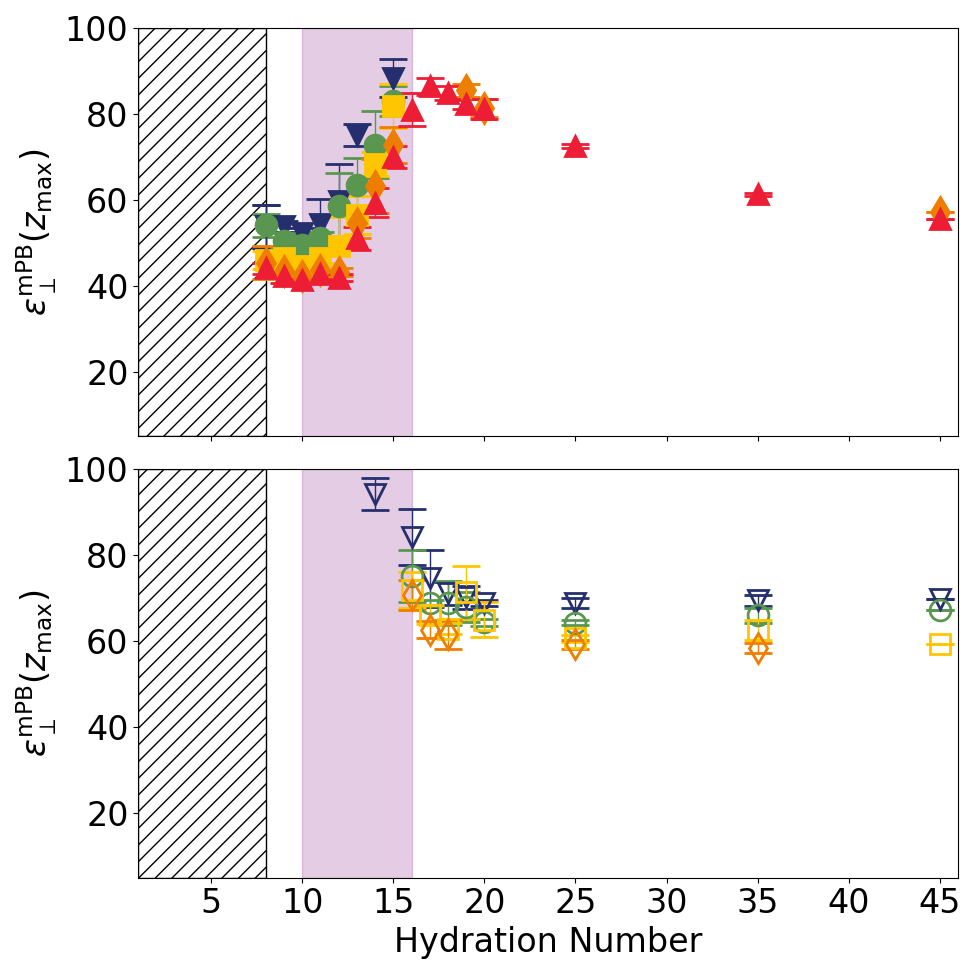}
   \caption{Maximum dielectric permittivity \textit{versus} hydration number at different temperatures, for bilayers in the fluid phase (top) and bilayers in the gel phase (bottom). The colours are the same as in Fig.~\ref{fig:ApL_vs_HN} and the open/closed markers have the same meaning. The purple region shows where we identify an attractive regime between the membranes (see Sect.~\ref{Subsec:Discussion:forces}). The mPB model cannot be used in the hatched region because the ions no longer follow a Boltzmann distribution.} 
   \label{fig:Epsilon_vs_HN}
\end{figure}

Once the mPB model has been solved in all systems, at all temperatures, one can extract the relative dielectric permittivities at the centre of the water channels by evaluating $\varepsilon_{\perp}^{\mathrm{mPB}}(z_{\mathrm{max}})$, where $z_{\mathrm{max}} = L_z / 2$ as shown in Fig.~\ref{fig:systemPresentation}. The resulting values are plotted in Fig.~\ref{fig:Epsilon_vs_HN} for bilayers in the fluid phase and in the gel phase.

The value at the centre of the water layer corresponds to the maximum of the dielectric permittivity profile. If one wanted to define an effective permittivity of the whole water slit, one would typically take the average of its inverse~\cite{bib:Borgis2023} $\varepsilon_{\perp}^{-1}$. Averaging over $\varepsilon_{\perp}^{-1}$ results in an effective permittivity that is dominated by the lower values of $\varepsilon_{\perp}$. Therefore, a typical effective permittivity of the whole water slit would be lower than $\varepsilon_{\perp}^{\mathrm{mPB}}(z_{\mathrm{max}})$. The value of such an effective average of the dielectric permittivity strongly depends on the definition of its thickness and position, which are not easily defined in the case of lipid bilayers. We therefore resorted to investigate the maximum of the permittivity profiles. 

\medskip

In the gel phase, the dielectric permittivity is rather constant upon dehydration while staying at high HN. However, it starts to raise at some point, which is counter-intuitive. The same phenomenon appears for the fluid phase, however in this case the raise seems to be continuous upon all the dehydration process. At the time we write this paper, we do not have a clear explanation for this increase.
On the one hand, it could be because the real dielectric permittivity is actually a non-local property, but it is described as a local one in the model. In other words, $\varepsilon_{\perp}^{\mathrm{mPB}}$ is an effective dielectric permittivity profile, optimized to describe the ionic charge densities, but not the real profile itself. It depends on the parametrized curve we chose to model it. Therefore, it is possible that this curve does not allow to describe the dielectric permittivity profile with enough accuracy and that the model compensates with odd values of the dielectric permittivity.
Noticeably, in the study by Borgis et al (Stockmayer fluid confined by Lennard-Jones planar walls~\cite{bib:Borgis2023}), this quantity also reaches a maximum well above the bulk value of the dielectric constant nearby the confining interface.

On the other hand, this increase could come from the mPB model itself. We brought two modifications to the standard PB model: we assumed a varying dielectric permittivity along the $z$ axis and added the Born energy in the Boltzmann distribution of the ions. The model may lack some other interactions in this complex environment.

\medskip

While the model gives high values of the dielectric permittivity at large hydration, we observe a strong decrease of the dielectric permittivity, down to a plateau value at HN~=~11 which is around half its value when the attraction arises. It is generally assumed that HN~=~10 is a threshold value, corresponding to the water molecules bound to the lipid heads~\cite{bib:Malik2021}. For this reason, the value of $\varepsilon_{\perp}^{\mathrm{mPB}}$ that remains close to 50 for the dehydrated systems (HN~<~12) can be interpreted as the dielectric permittivity in lipid head region, that becomes almost independent of the water content.

The plot of the maximum dielectric permittivity as a function of the interlayer water thickness is shown in Fig.~\ref{SI:fig:Epsilon_vs_dw} in the supplementary material.

\medskip

Under HN~=~8, the mPB model fails due to a lack of statistics. Because of the low number of anions and since the dynamics in the water channel between the membranes slows down a lot when reaching this dehydration state (even at high $T$), the charge density profile of the anions no longer follows a Boltzmann distribution.

\section{DISCUSSION}

\subsection{Validity of the mPB model}

\begin{figure}[bp]
   \centering
   \includegraphics[width=\columnwidth]{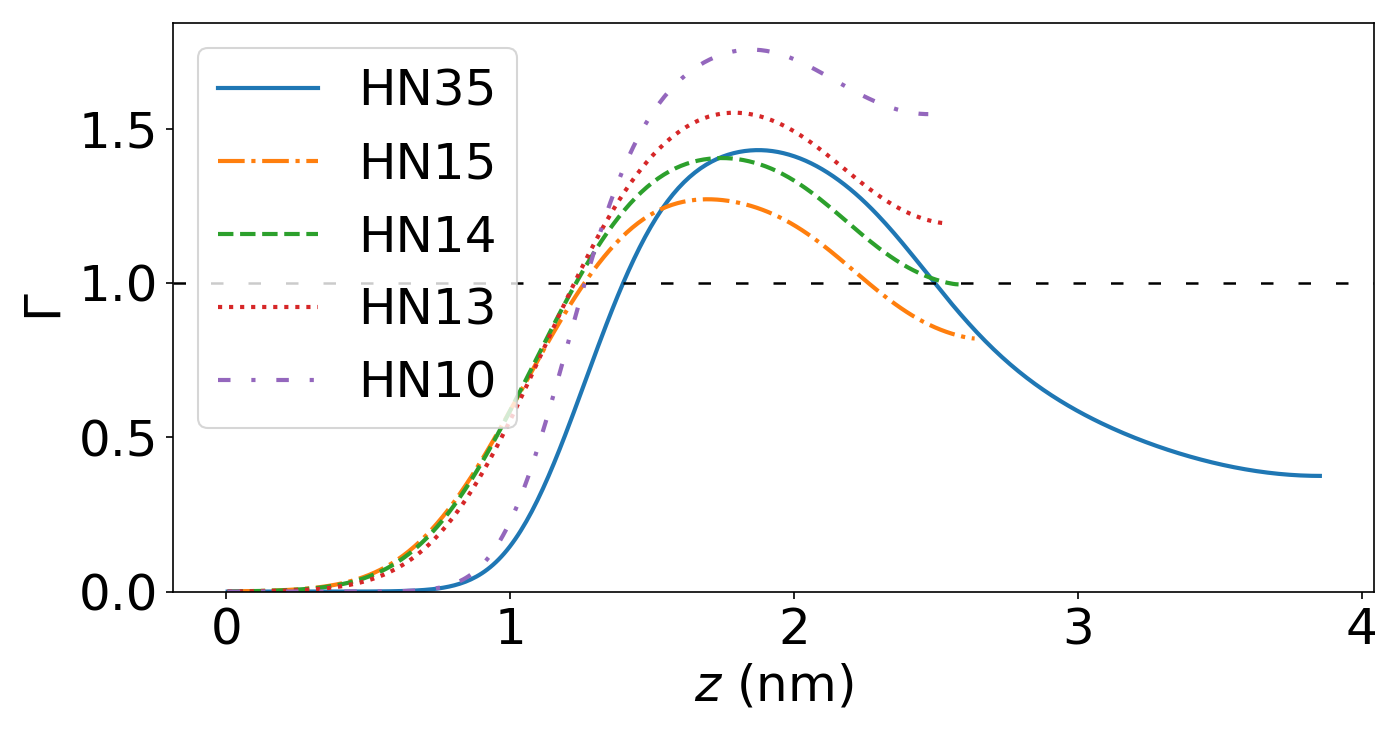}
   \caption{Evolution of the plasma parameter at different HN. Above the dashed line at $\Gamma$~=~1, ion correlations start to become non negligible.}
   \label{fig:PlasmaParameter}
\end{figure}

It is known that PB theory is less accurate when approaching SC regime. To estimate when such electrostatic couplings may appear locally, one can compute the plasma parameter $\Gamma$, which compares the electrostatic interaction energy between two ions and the thermal energy. It is given by~\cite{bib:Herrero2024}:
\begin{equation} \label{eq:PlasmaParameter}
    \Gamma(z) = \frac{\ell_{\mathrm{B}}(z)}{d_{\mathrm{ion}}(z)} \, ,
\end{equation}
where $d_{\mathrm{ion}}(z)$ is the typical distance between two ions and can be computed as follows: $d_{\mathrm{ion}}(z) = \sqrt[3]{C_{+}(z) + C_{-}(z)}$.
The value $\Gamma = 1$ is the threshold at which correlations between the ions start to become non negligible~\cite{bib:Herrero2024, bib:Levin2002}. Since these are not considered in the PB model, its accuracy decreases when $\Gamma$ exceeds this value.

\medskip

Fig.~\ref{fig:PlasmaParameter} shows the evolution of the plasma parameter in systems at $T$~=~353.15\,K, at different HN. At high hydration, the plasma parameter exceeds 1 in the lipid heads, but quickly decays to a lower value in the water. At HN~=~14 and below, $\Gamma$ stays over 1 in the water.
Because of the correlations, the results given by the mPB model in this hydration range will be less accurate, though they can still be analysed qualitatively. In addition, the plasma parameter could be underestimated since it is computed using the dielectric permittivity given by the mPB model.

\medskip

From their experimental study of DPPS membranes, Mukhina and al.~\cite{bib:Mukhina2019} have made an estimation of the dielectric permittivity of the confined interlayer water. They found that values between 10 and 30 gave a coupling constant $\Xi$ high enough to justify an attraction behaviour between the membrane in their system, within the SC theory. Their estimation of the dielectric constant took into account a measure of the interlayer water thickness, which corresponds to a HN between 8 and 10. The mPB model gives higher values for the dielectric permittivity at this hydration state, which could be caused by the discussed inaccuracies.

\subsection{Comparison with alternative ways of estimating the relative dielectric permittivity}

In this section, we compare the results obtained from the mPB model  described in this work with alternative ways of estimating the dielectric permittivity.

\medskip

\textit{Born dielectric permittivity}--- By getting rid of the electrostatic energy term the same way we did in Eq.~\eqref{eq:ionsConcentrationProduct}, i.e. by multiplying the ionic concentrations, one can compute a dielectric permittivity profile $\varepsilon_{\perp}^{\mathrm{Born}}$ which would be the effective permittivity in the Born model. This "Born dielectric permittivity" derives from Eq.~\eqref{eq:BornEnergy} and is given by:
\begin{equation} \label{eq:BornDielectricPermittivity}
    \begin{split}
        \frac{1}{\varepsilon_{\perp}^{\mathrm{Born}}(z)} & = \frac{1}{\varepsilon_{\perp}^{\mathrm{ref}}} + \frac{16 \pi \varepsilon_0}{\beta e^2} \frac{r_{+}r_{-}}{r_{+}+r_{-}} \\
        & \hphantom{{}= \frac{1}{\varepsilon_r^{\mathrm{ref}}} + \, } \times \ln \left( \frac{C_{\mathrm{salt}}}{\sqrt{C_{+}(z) \times C_{-}(z)}} \right) \, .
    \end{split}
\end{equation}

\begin{figure}[bp]
   \centering
   \includegraphics[width=\columnwidth]{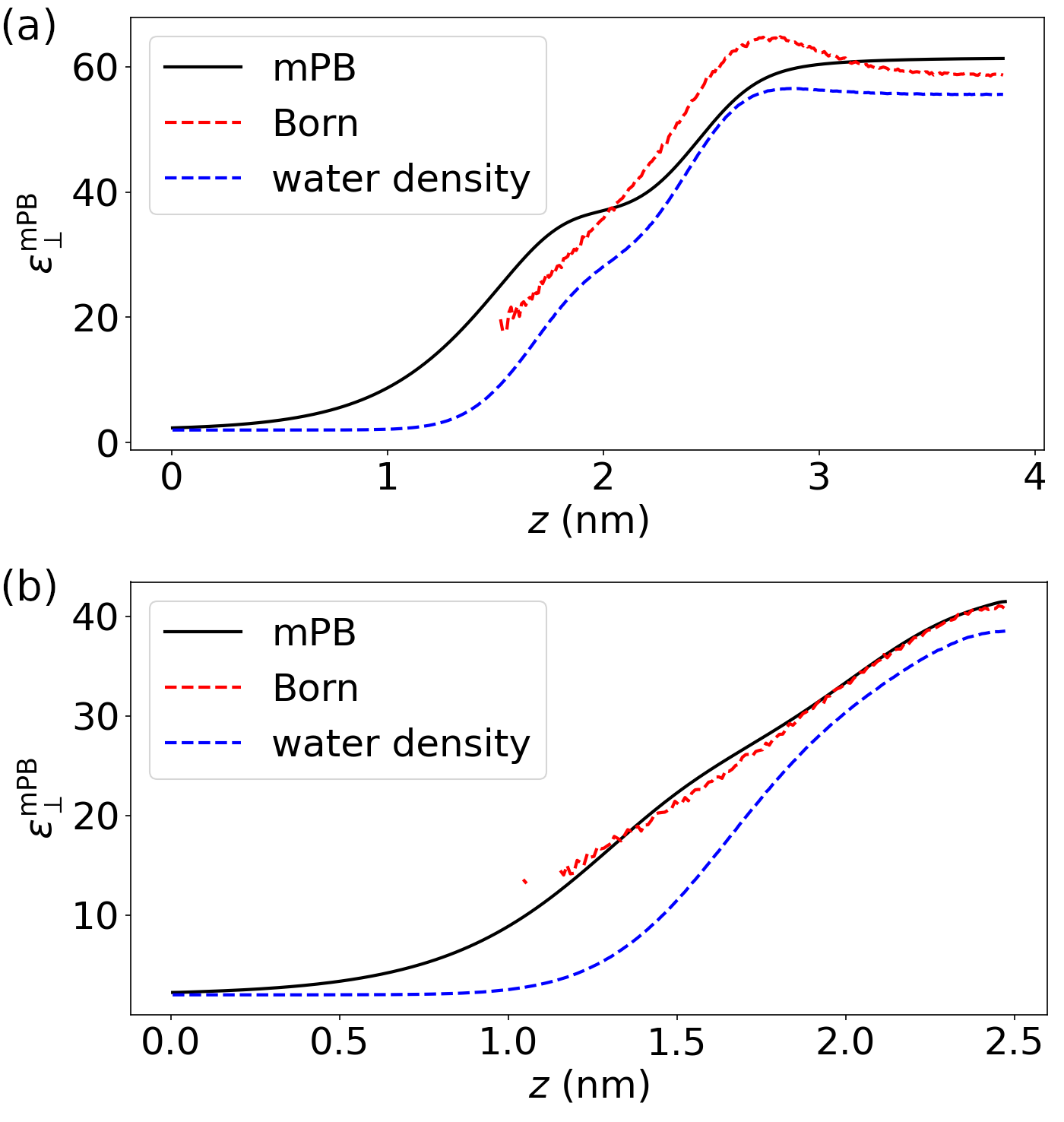}
   \caption{Comparison between $\varepsilon_{\perp}^{\mathrm{mPB}}$, $\varepsilon_{\perp}^{\mathrm{Born}}$ and $\varepsilon_{\perp}^{\mathrm{water~density}}$ for two systems at \mbox{$T$~=~353.15\,K}. (a) \mbox{HN~=~35}. (b) \mbox{HN~=~10}.}
   \label{fig:comparisonWithEpsilonBorn}
\end{figure}

Since the mPB model presented in this work highly relies on the Born model to compute the effective permittivity profile, comparing our results with the profile of $\varepsilon_{\perp}^{\mathrm{Born}}$ can give an insight on the self-consistency of the mPB model. Fig.~\ref{fig:comparisonWithEpsilonBorn} shows a comparison between $\varepsilon_{\perp}^{\mathrm{mPB}}$ and $\varepsilon_{\perp}^{\mathrm{Born}}$. Eq.~\eqref{eq:BornDielectricPermittivity} is not valid when the ionic concentrations are null, which is why $\varepsilon_{\perp}^{\mathrm{Born}}$ has no value in the tails region. Elsewhere however, it appears that the two profiles are quite similar, and even more at low HN.
Noticeably, in the mPB model, we have neglected the change in steric interactions between the water and lipid region, i.e. the variation of the free energy to form a cavity for the ions in the different locations of the system. The consistency between the profiles of $\varepsilon_{\perp}^{\mathrm{Born}}(z)$ and $\varepsilon_{\perp}^{\mathrm{mPB}}(z)$ illustrated in Fig.~\ref{fig:comparisonWithEpsilonBorn} supports this assumption.

\medskip

\begin{figure}[tbp]
   \centering
   \includegraphics[width=\columnwidth]{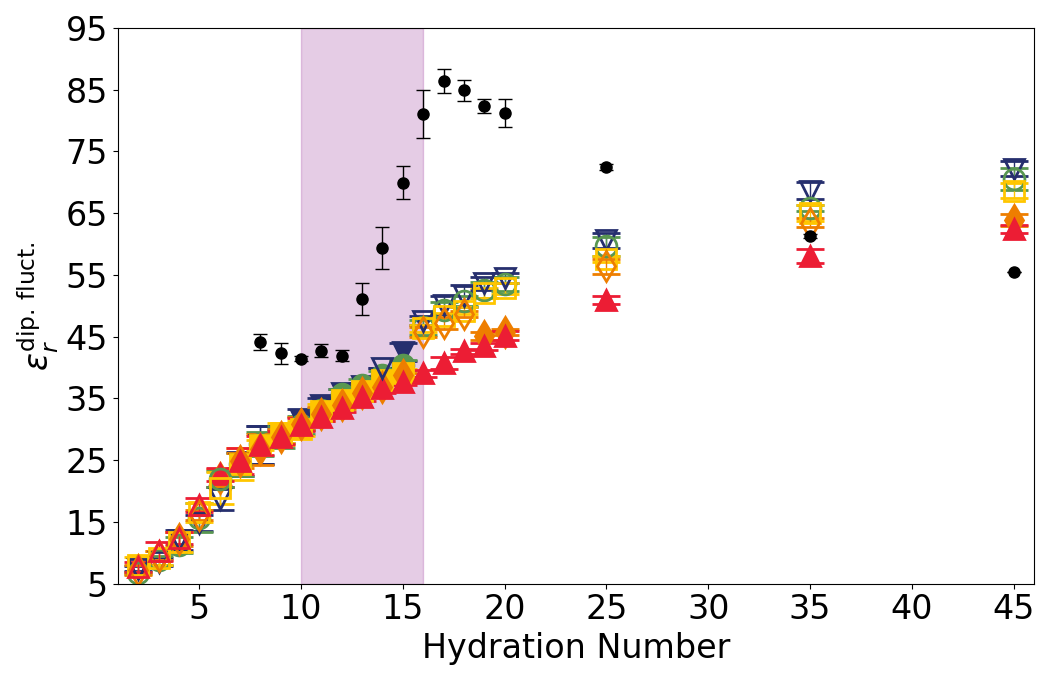}
   \caption{Dielectric response of the water molecules \textit{versus} hydration number at different temperatures. Blue dots correspond to $\varepsilon_{\perp}^{\mathrm{mPB}}(z_{\mathrm{max}})$ at 353.15\,K. The colours are the same as in Fig.~\ref{fig:ApL_vs_HN} and the open/closed markers have the same meaning. The purple region shows where we identify an attractive regime between the membranes (see Sect.~\ref{Subsec:Discussion:forces}).} 
   \label{fig:EpsilonFromDipoles_vs_HN}
\end{figure}

\textit{Fluctuation-dissipation Approach}--- 
The dielectric permittivity is commonly  calculated using the fluctuation-dissipation theorem that links the fluctuations of the total dipole moment $\bm{M}$ to the dielectric permittivity~\cite{bib:Kirkwood1939}. In a homogeneous medium containing solely water dipoles, the relation is:
\begin{equation} \label{eq:DipoleFluctuationDielectricPermittivity}
    \varepsilon^{\mathrm{dip.\,fluct.}} = 1 + \frac{\left< M^2 \right> - \left< M \right>^2}{3 \varepsilon_0 n_{\mathrm{w}} V_{\mathrm{w}} k_{\mathrm{B}} T} \, ,
\end{equation}
where $n_{\mathrm{w}}$ is the number of water molecules. Here we propose to estimate the dielectric permittivity through this formula, beyond its validity domain, since the system is inhomogeneous, and contains other dipoles than the water molecules. In practice, this approach should be considered as a way to assess the total water dipole fluctuations in a normalized representation.

\medskip

Results of this method are shown in Fig.~\ref{fig:EpsilonFromDipoles_vs_HN}. The plot of $\varepsilon^{\mathrm{dip.\,fluct.}}$ as a function of the interlayer water thickness is shown in Fig.~\ref{SI:fig:EpsilonFromDipoles_vs_dw} in the supplementary material.
At high hydration (HN~$\ge$~35), both mPB model and this method give similar results. However, they differ more and more upon dehydration, because of two main reasons. The first one is that $\varepsilon^{\mathrm{dip.\,fluct.}}$ only takes into account the water molecules, while $\varepsilon_{\perp}^{\mathrm{mPB}}$ acts as an effective permittivity in the system, and does not strictly corresponds to the dielectric permittivity of the sole water molecules. The second reason is that by essence, $\varepsilon^{\mathrm{dip.\,fluct.}}$ is monotonically decreasing upon dehydration, as is is linked to the decrease of the rotational degree of freedom of water molecules. At very low HN, the only remaining water molecules are surrounded by charged components (being either the ions or the lipid heads). The dipole moment of the water molecules is then constrained by these interactions and cannot really fluctuate, explaining the extremely low values of the permittivity obtained with this method in such regime.

\medskip

\textit{Correlation between permittivity and water density} --- 

Finally, the dielectric permittivity profile shapes shown in Figs.~\ref{fig:mPB_353K}(a) and ~\ref{fig:mPB_353K}(d) suggest  that the permittivity profile follows the density profiles of water, $\rho_{\text{w}}(z)$. Such an ad-hoc scaling was already proposed by M.\,Miettinen~\cite{bib:Miettinen2010} for cationic bilayers~\cite{bib:Gurtovenko2005, bib:Miettinen2009}. We model here this correlation with a function that interpolates between the two extreme values 
$\varepsilon_{\text{tails}}$ to  
$\varepsilon_{\perp}^{\mathrm{ref}}$ with a variation proportional to $\rho_{\text{w}}/\rho_{\text{w}}^{\text{bulk}}$:
\begin{equation} \label{eq:epsilon_water_density}
   \varepsilon_{\perp}^{\text{density}}(z) = \varepsilon_{\text{tails}} + \frac{\rho_{\text{w}}(z)}{\rho_{\text{w}}^{\text{bulk}}} \left[ \varepsilon_{\perp}^{\mathrm{ref}} - \varepsilon_{\text{tails}} \right].
\end{equation}
The values for $\varepsilon_{\perp}^{\mathrm{ref}}$ are reported in Tab.~\ref{SI:tab:calibrationValues}. 
The $\rho_{\text{w}}^{\text{bulk}}$ were chosen as the water densities at $z_{\mathrm{max}}$ for HN~=~45.

The comparison of $\varepsilon_{\perp}^{\text{density}}$  profile with the one of $\varepsilon_{\perp}^{\mathrm{mPB}}$ for two different HN at \mbox{$T$~=~353.15\,K} are reported in Fig.~\ref{fig:comparisonWithEpsilonBorn}.
It turns out that the profiles given by Eq.~\ref{eq:epsilon_water_density} are qualitatively similar. A first difference is that $\varepsilon_{\perp}^{\text{density}}(z_{\mathrm{max}})$ cannot describe the increase of $\varepsilon_{\perp}^{\mathrm{mPB}} (z_{\mathrm{max}})$ upon dehydration since its maximum remains $\varepsilon_{\perp}^{\mathrm{ref}}$ whatever the hydration level.
Moreover, Eq.~\ref{eq:epsilon_water_density}  underestimates both $\varepsilon_{\perp}^{\mathrm{mPB}}$ and $\varepsilon_{\perp}^{\mathrm{Born}}$. Especially in the region where the water density is very low, the contribution of the lipid dipoles to the permittivity $\varepsilon_{\perp}^{\mathrm{mPB}}$ is not negligible.

\subsection{Resulting force between the bilayers} \label{Subsec:Discussion:forces}

In the case of charged lipid bilayers with monovalent salt, the force between the bilayers is generally expected to be repulsive~\cite{bib:Marra1986}, but recent observations of DPPS membranes with NaCl salt suggest that this interaction could also be attractive~\cite{bib:Mukhina2019}. This attractive behaviour was associated to a decrease of permittivity in the water layer.

\medskip

In this context, we also aimed at determining if the membranes in our simulations were attracting or repelling each others. The force between lipid bilayers can be extracted from MD simulations, for example by calculating the water chemical potential~\cite{bib:Schneck2012, bib:Schlaich2024}.
It would be very interesting to perform such calculations in the present case, but it is beyond the scope of the present article. Given the relatively high computational cost, we resorted to a different approach, valid for membrane simulations without surface tension~\cite{bib:Smirnova2013}. It involves looking at the Helmholtz free energy $F$, which depends on the volume $V$, the temperature $T$ and the total number of particles $N$ in the system. Smirnova and al. expressed, in the ($NVT$) ensemble, the excess of free energy $\Delta F$, with respect to a single fully hydrated (in bulk water) lipid bilayer with no surface tension, as follows:
\begin{equation} \label{eq:excessFreeEnergySmirnova}
    \Delta F = \Delta F_{\mathrm{bilayer}} \, + \Delta F_{\mathrm{water}} \, + g(d_{\mathrm{w}})A \, ,
\end{equation}
where $\Delta F_{\mathrm{bilayer}}$ is the excess free energy contribution coming from the elastic deformation of the bilayer, $\Delta F_{\mathrm{water}}$ is the one coming from the compression of the bulk of water surrounding the membrane, $A$ is the surface of the interface and $g$ is the interface potential.
In stacked bilayers systems, the resulting force between lipid bilayers is generally considered as the sum of three main contributions. The first one comes from the van der Waals interactions, which are attractive. The second one, the hydration repulsion, prevents the membranes from sticking to each others: it is a short-ranged interaction and decays exponentially with the thickness of the interlayer confined water~\cite{bib:Xu2010}. The third one comes from the electrostatic interactions.
We suppose here that these interactions between membranes can be described within the last term $g(d_{\mathrm{w}})A$, i.e. proportional to the membrane area, and as a function of the water thickness $d_{\mathrm{w}}$. This assumption turned out to be relevant for the neutral lipid membranes~\cite{bib:Smirnova2013}, and we assume it to hold for charged membranes as well.
The potential $g$ depends on the thickness of the water thickness $d_{\mathrm{w}}$, which can be estimated from the HN as:
\begin{equation} \label{eq:waterThicknessFromHN}
    d_{\mathrm{w}}(\mathrm{HN}) = \frac{2 \times \mathrm{HN} \times V_{\mathrm{w}}}{A_{\mathrm{lipid}}} \, ,
\end{equation}
where $V_{\mathrm{w}} = 29.7$\,\angstrom\textsuperscript{3} is the volume of a water molecule and $A_{\mathrm{lipid}}$ is the area per lipid. We did not consider the dependence of this molecular volume on the temperature here as we were not interested in quantitative measurements but rather in the qualitative behaviour in the simulation.

\medskip

Smirnova and al. stated that since the variation of the simulation box volume is negligible, the variations of Gibbs free enthalpy $G$ are similar to the Helmhotz free energy variations : $\Delta G \approx \Delta F$. They were then able to express the repulsion pressure $P(d_{\mathrm{w}}) = - g'(d_w)$, the derivative of the interface potential $g$ can be expressed as a function of the area variation \mbox{$\Delta A = A-A_0$}:
\begin{equation} \label{eq:repulsionPressure}
    P \left( d_{\mathrm{w}} \right) = - \frac{K_{\mathrm{A}}}{A_0}  
    \left[
    \frac{\Delta A(d_{\mathrm{w}})}{d_{\mathrm{w}}} 
    + 
    \int_{d_{\mathrm{w}}}^{\infty} 
    \frac{\Delta A(d)}{d^2} 
    \mathrm{d}d
    \right],
\end{equation}
where $K_{\mathrm{A}}$ is the area compressibility modulus, $A_0$ is the reference area at the largest HN, and $A(d_{\mathrm{w}})$ is the area for a given water thickness, i.e. a given HN. The attractive vs repulsive behaviour between the membranes depends on the sign of this pressure. It will be repulsive if \mbox{$P(d_{\mathrm{w}}) > 0$} and attractive otherwise. Noticeably, in the work by Smirnova et al., Eq.~\ref{eq:repulsionPressure} was used under several  conditions~\cite{bib:Smirnova2013}: (1) there is no fluid/gel phase change in the bilayer. (2) $K_A$ does not vary upon dehydration. (3) the integral in Eq.~\ref{eq:repulsionPressure} is supposed to be negligible and the sign of the area variation \mbox{$\Delta A(d_{\mathrm{w}})$} directly determines the sign of the pressure $P(d_{\mathrm{w}})$. Since $K_A$, $d_{\mathrm w}$ and $A_0$ are positive, a larger area denotes an attractive behaviour while a smaller area stands for a repulsion.

To constrain our analysis within the condition (1), we shall solely discuss here simulations in the fluid phase. Concerning the condition (2), Fig.~\ref{SI:fig:compressibilityModulus_vs_HN} shows that for the simulations in the fluid phase, the compressibility modulus is practically constant in the range \mbox{$11 \leq$ HN $\leq 45$}. Concerning the condition (3), Fig.~\ref{SI:fig:P1_P2_vs_dw} comparing the two terms of Eq.~\ref{eq:repulsionPressure} shows that the second term is negligible at larger HN, but it can become non negligible at the lowest HNs. Nevertheless, in the range of hydration where $K_A$ is constant, at larger distances $d$, $\Delta A(d)$ does not change sign. In this hydration range, the two terms of Eq.~\ref{eq:repulsionPressure} are of the same sign. 
Therefore, we analyse in the following the sign of $\Delta A(d_{\mathrm{w}})$ to infer the sign of $P(d_{\mathrm{w}})$.

\medskip

\begin{figure}[tbp]
   \centering
   \includegraphics[width=\columnwidth]{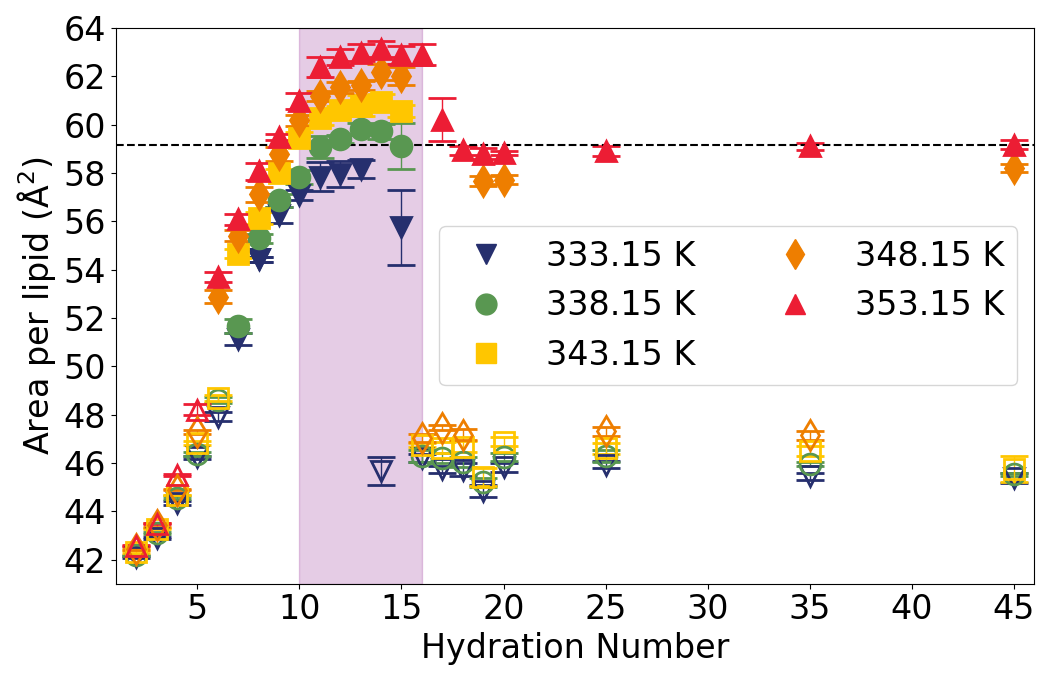}
   \caption{Area per lipid \textit{versus} hydration number at different temperatures. Filled markers represent bilayers in the fluid phase and open markers represent bilayers in the gel phase. The blue dotted line shows the reference area per lipid $A_0$ at $T = 353.15$\,K. At this temperature, following Eq.~\eqref{eq:repulsionPressure}, we identify the purple region as the hydration range at which the attractive behaviour appears.}
   \label{fig:ApL_vs_HN}
\end{figure}

Fig.~\ref{fig:ApL_vs_HN} shows the area per lipid as a function of the HN, at the five studied temperatures. The same plot as a function of the interlayer water thickness $d_{{\mathrm w}}$ is shown in Fig.~\ref{SI:fig:ApL_vs_dw} in the Supplementary Material. Since the number of lipids per layer remains constant, it is equivalent to study the total area variation or the area per lipid variation. It appears that at high HN (typically above 20), there is no significant change in the area per lipid upon dehydration. This indicates that in this HN range the bilayers are fully hydrated and they do not interact significantly. However, if we keep dehydrating the system, two thresholds appear. The first one is between HN~=~20 and HN~=~15. After it is reached, we see an increase in the area per lipid. The second threshold appears around HN~=~10-11, below which the opposite phenomenon occurs: the area per lipid starts to decrease. These changes in the area per lipid can even induce phase transitions in the bilayers. In the coldest systems between HN~=~20 and HN~=~15, the bilayer undergoes a transition from the gel phase to the fluid phase. At lower HN, it undergoes a transition from the fluid phase to the gel phase in all the simulations.

\medskip

One has to take care of these phase changes when interpreting the differences in area per lipid using Eq.~\eqref{eq:repulsionPressure}. In this work, we safely focus on the system at 353.15\,K, for which the bilayers stay in a fluid phase down to HN~=~6. At high hydration, the system is in an equilibrium state and there is no significant force acting on the bilayers.
However, after the first threshold is met, the area per lipid is higher than the reference value. This corresponds to a negative value for the repulsion pressure, i.e. an attractive interaction between the bilayers.
If the dehydration continues after the second threshold, the conditions necessary for Eq.~\ref{eq:repulsionPressure} do not hold any longer in our simulations. But the strong decrease in area might be the hint that the short-range hydration repulsion well known in the literature for any kind of lipid bilayers~\cite{bib:Xu2010} becomes more important.

To conclude, even if we do not calculate explicitly the interface potential, the area per lipid variations indicate some inter-membrane attraction at \mbox{$T$~=~353.25\,K} in the range of \mbox{11 $\leq$ HN $\leq$ 16}, which is the same HN-range where the permittivity of the water layer drops.

%% file: 04_conclusion.tex
\section{CONCLUSION}

We have presented a 1D modified Poisson-Boltzmann model, which allows the dielectric permittivity to vary along the $z$ axis and takes into account the solvation energy of the ions (Born energy) in their Boltzmann distribution. 

\medskip

In the model's equation, the two free charge densities come from the ions and from the lipids. The first one has an analytical description (a Boltzmann distribution) while the second one is added as an external fixed contribution obtained from MD simulations. Two unknowns remain in the equation: the electric potential and the dielectric permittivity. We left the electric potential as the unknown of the equation and we made an educated guess on the shape of the dielectric permittivity profile. With the help of a previous work that could only compute this profile in systems without free charges~\cite{bib:Schlaich2016}, we modelled the dielectric permittivity profile as a parametrized curve and optimized the parameters so that the analytical description of the ions in the model fitted at best the charge densities obtained from MD simulations.

\medskip

Since this model can estimate the relative dielectric permittivity in a system containing free charges, we were able to use it to investigate water confined between charged lipid bilayers. The results show a self-consistence of the model, giving a dielectric permittivity profile close from the one that directly derives from the Born model. Moreover, when focusing on the dielectric permittivity at the centre of the interlayer channel, the mPB model shows a strong decrease of the dielectric constant when entering a hydration regime where the membranes attract each others. This observation is in agreement with the common interpretation of the phenomenon in the literature. At intermediate hydration levels, the mPB model results in a local maximum of the dielectric constant that is higher than the bulk value. This may be due to the fact that our profiles are by construction smoothed, coarse-grained, and treat the dielectric permittivity as a local property. At high hydration, it is in good agreement with a classic way of computing the dielectric constant from the fluctuations of the water molecules dipole moment. However upon dehydration, it diverges rather quickly from this method, which only considers the water molecules.

\medskip

In the end, the model presented in this work is generic and thus can be applied in a large variety of systems where an electrolyte solution is confined between charged walls. We expect the mPB model to be more adapted to the simpler case of homogeneous planar walls, as the thermal fluctuations of the membranes are not taken into account in the model, since the lipid charge density profile is averaged over the simulation length. The curve to adjust the inverse permittivity profile would have to be adapted to these new systems.
This approach could also be useful to estimate the dielectric permittivity profile in biological systems, where the biological fluid of interest is not too much confined so the distributions of the salt it contains can be modelled with a Boltzmann distribution.

%% file: 05_thanks.tex
\section*{SUPPLEMENTARY MATERIAL}
The supplementary material includes (I) details on the numerical solving of the mPB model, (II) the inverse dielectric permittivity profile for a DPPC bilayer, (III) inverse dielectric permittivity profiles for the DPPS bilayers, and their variation when the fitting function includes 0, 1 or 2 Gaussians, (IV) details on the calibration of the mPB model and the independent permittivity measurements on bulk water, (V) a comparison of Born and electrostatic energy profiles, (VI) the area compressibility modulus as a function of hydration level at different temperatures, (VII) the dependence of the area per lipid,  and relevant permittivities as a function of $d_{\textrm{w}}$, and (VII) the list of simulated systems and their compositions.

\section*{ACKNOWLEDGMENTS}

We thank Hélène Berthoumieux, Ananya Debnath, Marcus Müller and Alexander Schlaich for fruitful discussions.
This study was supported by the French Agence Nationale de la Recherche through the Banana\_Slip project (ANR-21-CE30-0026).
We used computer and storage resources by GENCI at IDRIS thanks to the grants A0110807662 and A0150807662 on the supercomputer Jean Zay's V100 partition.
The authors thank the reviewers for their critics and insightful suggestions.

\section*{AUTHOR DECLARATIONS}

\textit{Conflict of interest} --- The authors have no conflict to disclose.

\section*{DATA AVAILABILITY}

The data that support the findings of this study are available from the corresponding author upon reasonable request. The complete trajectories, some input and output files for  HN~=~35 and HN~=~10  at $T=353.15$\,K have been published on Zenodo repository~\cite{zenodo:gardre_2025_16280641}, to permit a reference in MDverse~\cite{bib:Tieman2024} and NMRlipids~\cite{bib:Kiirikki2024} catalogs. Other trajectories are available upon reasonable request to the authors.

%% file: 06_references.tex
\phantomsection\addcontentsline{toc}{section}{REFERENCES}

\bibliographystyle{unsrt}
\bibliography{references_mainText}

%% file: suppMat.tex
\section{Numerical solving of the mPB model} \label{SI:sec:numericalSolving}

A custom Python code has been written to solve the mPB model. It is available here: \url{https://github.com/lgardre/mPB.git}. Let us develop Eq.~\eqref{eq:modifiedPoisson1D}, using Eqs.~\eqref{eq:BornEnergy} and \eqref{eq:modifiedIonsBoltzmannDistribution}:
\begin{equation} \label{SI:eq:modifiedPoisson1D_developped}
    \begin{split}
        \frac{\mathrm{d}^2 \phi}{\mathrm{d}z^2} & = \frac{\beta e^2}{\varepsilon_0 \varepsilon_{\perp}(z)} \biggl\{ C_{\mathrm{salt}} \Bigl( \e^{- \beta W_{-}(z) + \phi(z)} - \e^{- \beta W_{+}(z) - \phi(z)} \Bigr) + n_{\mathrm{lipids}}(z) \biggr\} - \frac{1}{\varepsilon_{\perp}(z)} \frac{\mathrm{d}\varepsilon_{\perp}(z)}{\mathrm{d}z} \frac{\mathrm{d}\phi(z)}{\mathrm{d}z} \, ,
    \end{split}
\end{equation}
with $\rho_{\mathrm{lipids}}(z) = -e \, n_{\mathrm{lipids}}(z)$.

\medskip

To solve this equation, we use the \texttt{solve\_bvp()} function from \texttt{scipy.integrate}. However, this function can only solve first order ODEs. Let us then rewrite Eq.~\eqref{eq:modifiedPoisson1D} as a system of first order ODEs. We define $\phi_1(z)$ and $\phi_2(z)$ as follows:
\begin{equation} \label{SI:eq:phi1}
\phi_1(z) = \phi(z) ~ \mathrm{and} ~ \phi_2(z) = \frac{\mathrm{d}\phi_1(z)}{\mathrm{d}z} = \frac{\mathrm{d}\phi(z)}{\mathrm{d}z} \, .
\end{equation}

\medskip

The system of equations then follows:
\begin{equation} \label{SI:eq:mPB_ODE_system}
    \left\{
    \begin{aligned}
      \frac{\mathrm{d} \phi_1(z)}{\mathrm{d}z} & = \phi_2(z)\\
      \frac{\mathrm{d} \phi_2(z)}{\mathrm{d}z} & = \frac{\beta e^2}{\varepsilon_0 \varepsilon_{\perp}(z)} \biggl\{ C_{\mathrm{salt}} \Bigl( \e^{- \beta W_{-}(z) + \phi_1(z)} - \e^{- \beta W_{+}(z) - \phi_1(z)} \Bigl) + n_{\mathrm{lipids}}(z) \biggr\} - \frac{1}{\varepsilon_{\perp}(z)} \frac{\mathrm{d}\varepsilon_{\perp}(z)}{\mathrm{d}z} \phi_2(z)
    \end{aligned}
  \right. \, .
\end{equation}

\medskip

We then introduce the inverse of the dielectric permittivity $\varepsilon_{\perp}^{-1}(z)$, so we rewrite Eq.~\eqref{SI:eq:mPB_ODE_system} accordingly:
\begin{equation} \label{SI:eq:mPB_ODE_system_with_inverse_permittivity}
    \left\{
    \begin{aligned}
      \frac{\mathrm{d} \phi_1(z)}{\mathrm{d}z} & = \phi_2(z)\\
      \frac{\mathrm{d} \phi_2(z)}{\mathrm{d}z} & = \frac{\beta e^2}{\varepsilon_0} \varepsilon_{\perp}^{-1}(z) \biggl\{ C_{\mathrm{salt}} \Bigl( \e^{- \beta W_{-}(z) + \phi_1(z)} - \e^{- \beta W_{+}(z) - \phi_1(z)} \Bigl) + n_{\mathrm{lipids}}(z) \biggr\} + \frac{1}{\varepsilon_{\perp}^{-1}(z)} \frac{\mathrm{d}\varepsilon_{\perp}^{-1}(z)}{\mathrm{d}z} \phi_2(z)
    \end{aligned}
  \right. \, .
\end{equation}

\medskip

In the code, the inverse of the dielectric permittivity is implemented as follows:
\begin{equation} \label{SI:eq:RealEpsilonPerpInvModel}
    \begin{split}
        \varepsilon_{\perp}^{-1} \left( z, z_1, K, z_2, z_3, C, \varepsilon_{\perp}^{\mathrm{tails}}, \varepsilon_{\perp}^{\mathrm{plateau}} \right) = & \, \frac{1}{\varepsilon_{\perp}^{\mathrm{plateau}}} \\
        & + \left( \frac{1}{\varepsilon_{\perp}^{\mathrm{tails}}} - \frac{1}{\varepsilon_{\perp}^{\mathrm{plateau}}} \right) \times \frac{1}{1 + \exp \Bigl[ K \left( z - z_1 \right) \Bigr]} \\
        & + \frac{C}{\sqrt{2 \pi} \frac{z_3 - z_2}{6}} \exp \left[ - \left( z - \frac{z_2 + z_3}{2} \right)^2 \times \frac{1}{2 \left( \frac{z_3 - z_2}{6} \right)} \right] \, .
    \end{split}
\end{equation}
In this equation, the parameters $z_2$ and $z_3$ are linked to $\sigma$ and $\mu$ from \eqref{eq:epsilonPerpInvModel} with the following relations:
\begin{equation} \label{SI:eq:linkSigmaMu_z2z3}
    \sigma = \frac{z_2 - z_3}{6} \, , \mu = \frac{z_2 + z_3}{2}\, . 
\end{equation}

\clearpage 

\section{Inverse dielectric permittivity profile for DPPC bilayer} 
\label{SI:sec:epsilonperpDPPC}

\begin{figure}[h!]
    \centering
\includegraphics[width=0.6\textwidth]{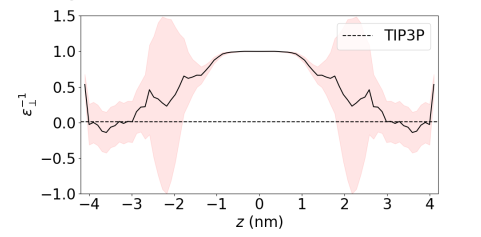}
    \caption{Perpendicular profiles of the dielectric permittivity in a
DPPC bilayer at \mbox{HN~=~45} and \mbox{$T$~=~333.15\,K}, calculated using MAICoS. The profiles are centred on the bilayer: lipid tails are at the centre, heads are around $z\simeq 2$ nm
and water is at the left and right boundaries. The dashed line gives the reference dielectric permittivity of TIP3P water at the same temperature ($\simeq 83$ degrees celcius). The red region corresponds to the uncertainty on the measure.}
    \label{SI:fig:epsilon_perp}
\end{figure}

Independant simulations were performed for CHARMM36-DPPC bilayers using the same protocol  and simulations parameters as for the DPPS described in the main text, but without adding ions in the solvant that is composed of pure mTIP3P water. The production runs were of 4~$\mu$s, with registration of the conformations every 0.1~ns. The inverse dielectric permittivity profile obtained using MAICos is displayed in Fig.~\ref{SI:fig:epsilon_perp}.

\clearpage 

\section{Inverse dielectric permittivity profiles and impact of Gaussians added to the sigmoid function} 
\label{SI:sec:gaussians}

\begin{figure}[h!]
   \centering
   \includegraphics[width=\textwidth]{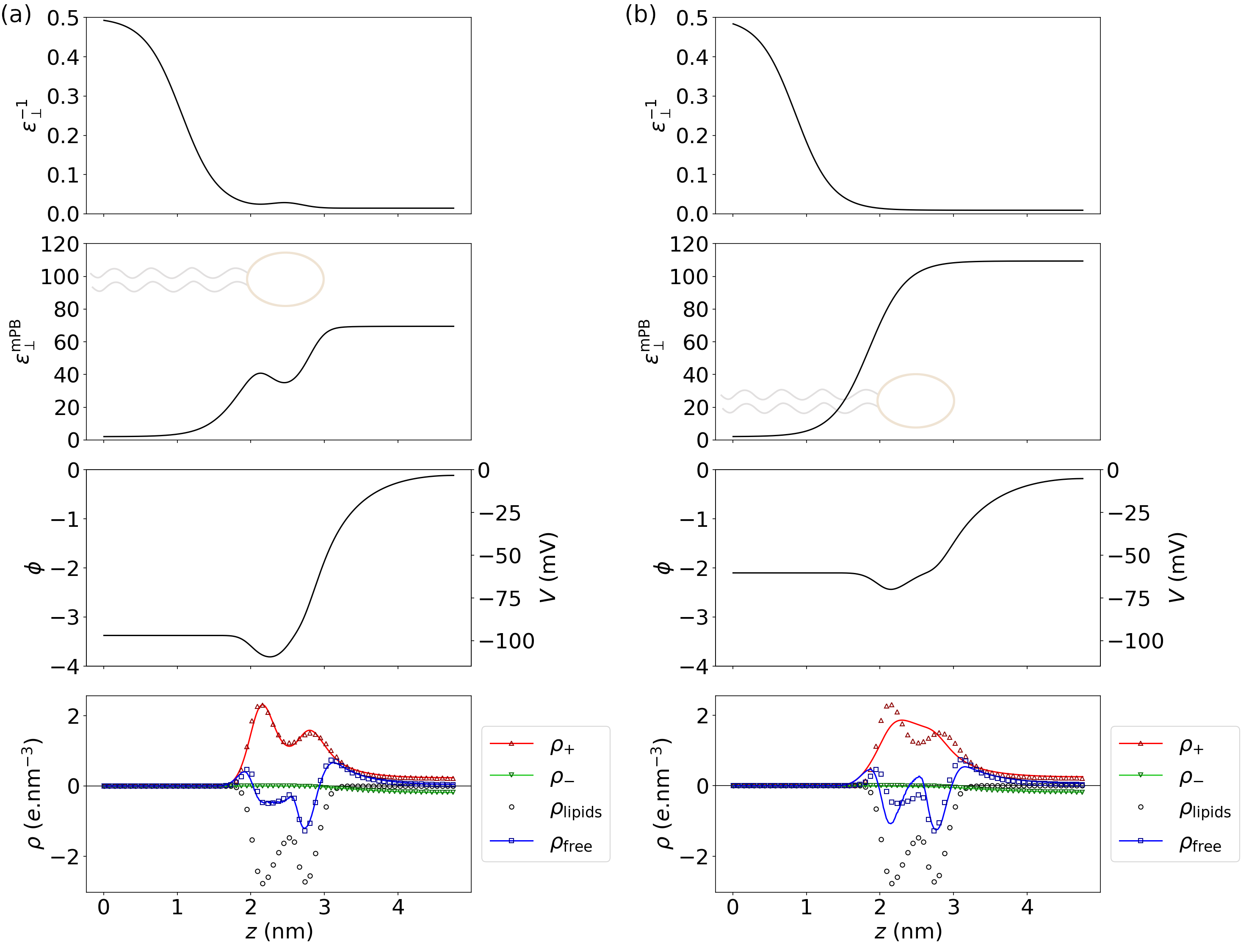}
   \caption{Results from the modified Poisson Boltzmann model at HN~=~35 and $T$~=~333.15\,K (lipid bilayer in the gel phase). The lipid in the background represents the approximative $z$ position of the tails and the heads in the lipid layer. (a) The inverse of the dielectric permittivity is modelled with a sigmoid and a Gaussian. (b) The inverse of the dielectric permittivity is only modelled with a sigmoid.}
   \label{SI:fig:mPB_333K}
\end{figure}

\begin{figure}[h!]
   \centering
   \includegraphics[width=0.75\textwidth]{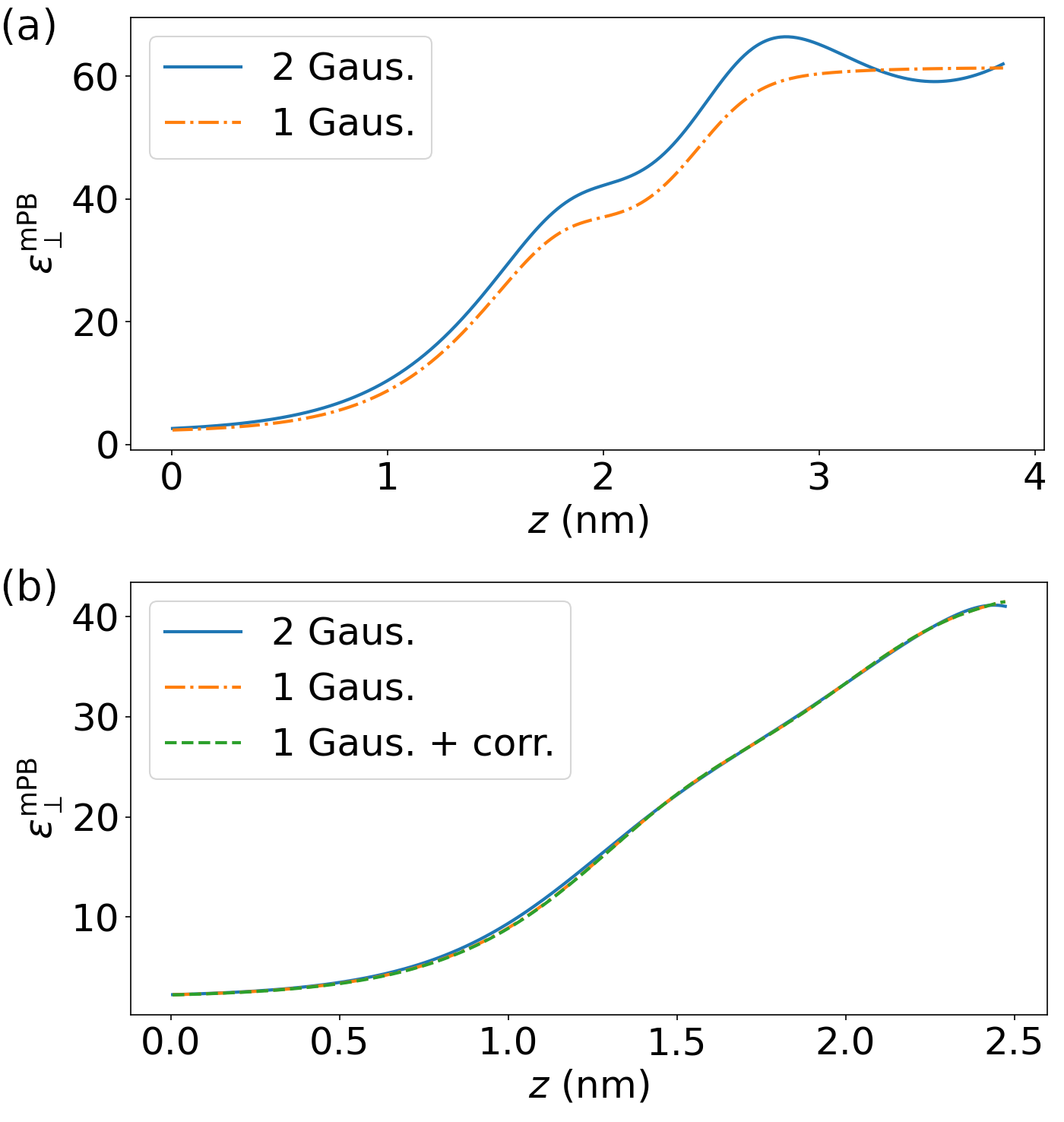}
   \caption{Permittivity profiles obtained from the modified Poisson Boltzmann model at $T$~=~353.15\,K (fluid phase). The inverse of the dielectric permittivity is modelled with a sigmoid plus one or two Gaussian functions. (a) HN~=~35. (b) HN~=~10. In this case, an additional small correction is added close to $z=z_{\mathrm{max}}$ to impose that the permittivity has no derivative at this point.}
   \label{SI:fig:mPB_353K_gaussians}
\end{figure}

\clearpage 

\section{Details about the calibration of the mPB model} \label{SI:sec:calibrationDetails}

\begin{figure}[h!]
   \centering
   \includegraphics[width=\textwidth]{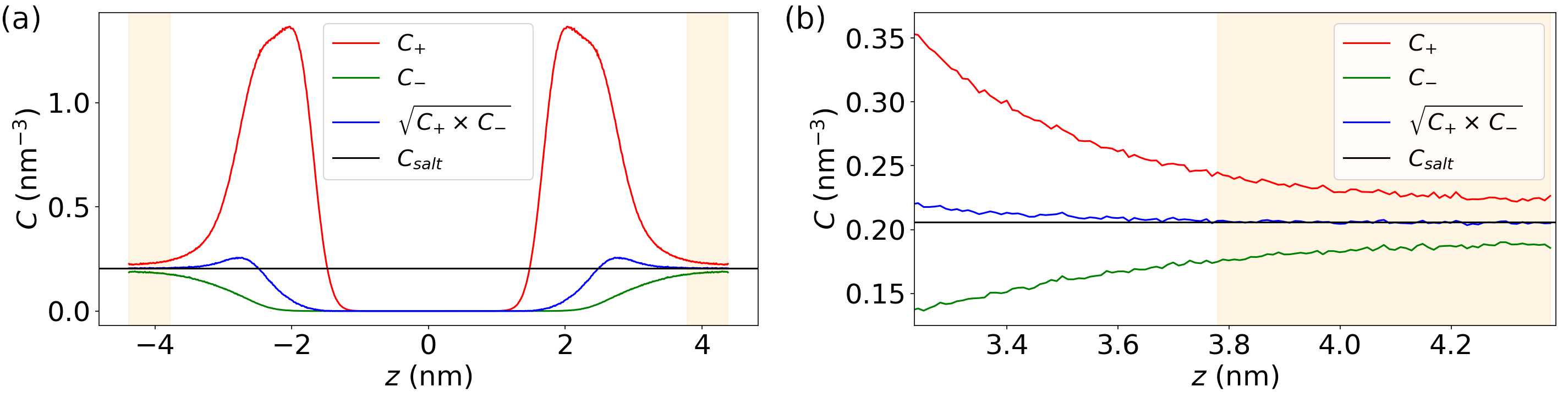}
   \caption{First step of the calibration on the system at HN~=~45 and $T$~=~353.15\,K. (a) Ionic concentrations in the system. The salt concentration, given by Eq.~\eqref{eq:CsaltCalibration}, is constant in the highlighted regions. (b) Zoom on a highlighted region. The constant $C_{\mathrm{salt}}$ is obtained by averaging the values of the salt concentration in both highlighted regions.}
   \label{SI:fig:mPB_calibration}
\end{figure}

\medskip

\begin{table}[h!]
    \centering
    \begin{tblr}{ | c | c | c || c | c | } 
        \hline
        $T$\,(K) & $C_{\mathrm{salt}}$\,(nm\textsuperscript{-3}) & $\varepsilon_{\perp}^{\mathrm{ref}}$ & $\varepsilon_{\mathrm{pure~water}}^{\mathrm{mTIP3P}}$ & $\varepsilon_{\mathrm{NaCl~0.25\,M}}^{\mathrm{mTIP3P}}$\\
        \hline
        333.15\,K & 0.187~$\pm$~0.002 & 69.82~$\pm$~1.69 & 84~$\pm$~6& 76~$\pm$~2 \\ 
        \hline[dotted]
        338.15\,K & 0.184~$\pm$~0.003 & 67.30~$\pm$~1.69 & 82~$\pm$~4 & 73~$\pm$~2 \\ 
        \hline[dotted]
        343.15\,K & 0.182~$\pm$~0.003 & 59.31~$\pm$~1.43 & 80~$\pm$~4 & 72~$\pm$~2 \\ 
        \hline[dotted]
        348.15\,K & 0.209~$\pm$~0.002 & 57.20~$\pm$~1.52 & 74~$\pm$~1 & 67~$\pm$~1 \\ 
        \hline[dotted]
        353.15\,K & 0.206~$\pm$~0.002 & 55.51~$\pm$~1.62 & 68~$\pm$~3 & 66~$\pm$~1 \\ 
        \hline
    \end{tblr}
    \caption{Values of the calibrated parameters in the mPB model, at each studied temperature. Data labelled $\varepsilon^{\mathrm{mTIP3P}}$ was computed by applying the Kirkwood relation in independent simulations containing either pure water or a 0.25\,M NaCl solution. The permittivity of pure bulk mTIP3P decreases by increasing temperature. Depending on temperature, the addition of 0.25 M NaCl diminishes the relative dielectric permittivity by about 3 to 11\%. }
    \label{SI:tab:calibrationValues}
\end{table}

The values for the mTIP3P model, either pure or with  0.25\,M NaCl were calculated in independant simulations. A cubic box of pure water or salted water of 5 nm size was contructed using CHARMM-GUI.
The system was then simulated in the NPT ensemble at 1 atm at the given temperature during 10~ns,  and the simulation was repeated 10 times using different seeds for the velocities. The parameters for the simulations and thermostat/barostat using Gromacs v.2024 were the same as the ones described in the main text for the lipidic systems.  The permittivity was calculated  using the \texttt{gmx dipoles} tool.

\begin{figure}[h!]
   \centering
   \includegraphics[width=0.75\textwidth]{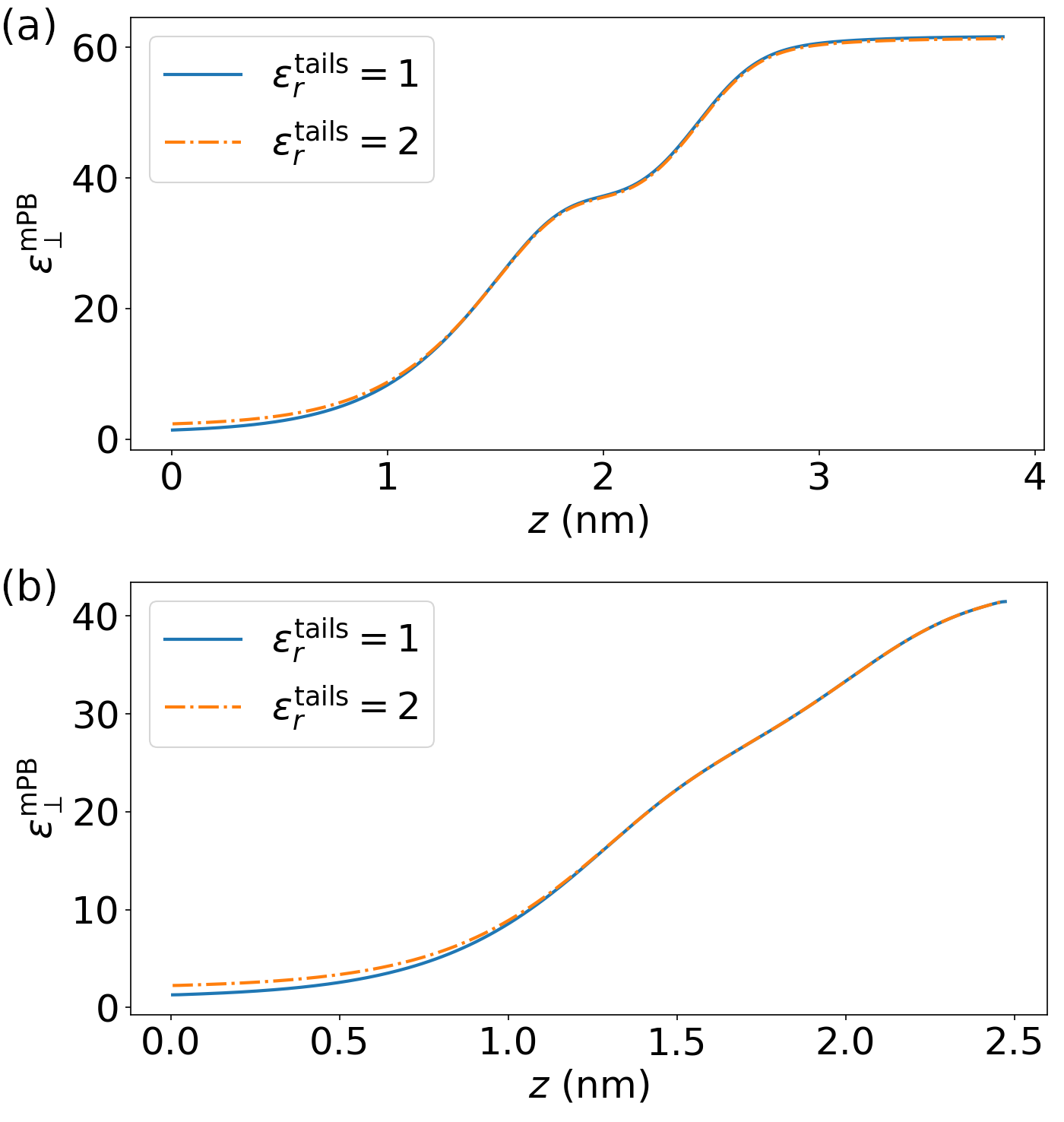}
   \caption{Dielectric permittivity profiles with the permittivity in the tails set to 1 or 2 for simulations at $T$~=~333.15\,K. \textbf{(a)}~HN35. \textbf{(a)}~HN10.}
   \label{SI:fig:epsilonR_tails}
\end{figure}

\clearpage 

\section{Repulsion at the interface due to the Born energy}
\label{SI:sec:energyComparison}

\begin{figure}[h!]
   \centering
   \includegraphics[width=0.6\columnwidth]{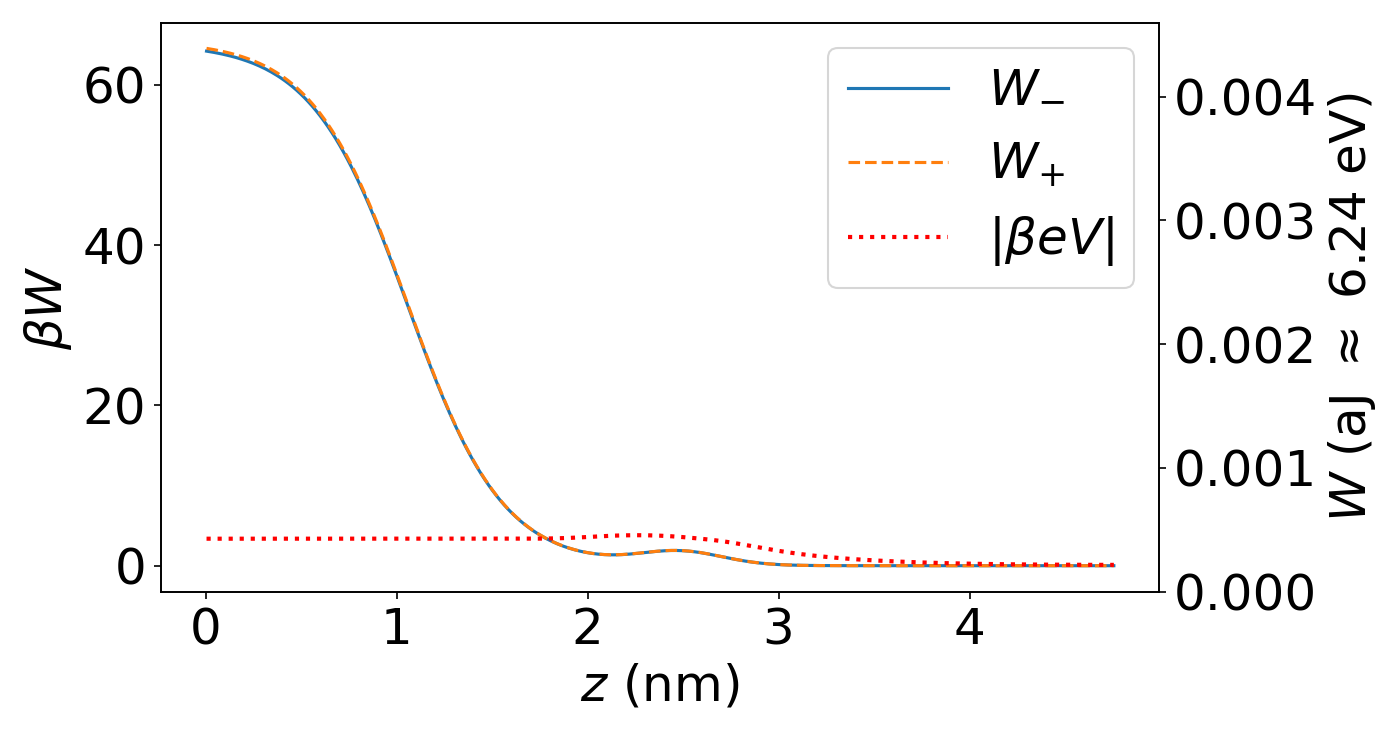}
   \caption{Comparison of the reduced Born energy profiles $\beta W_{\pm}(z)$ for cations and anions, and the absolute value of the reduced electric potential $\beta e V(z)$, for the system at HN~=~35 and $T$~=~333.15\,K. The energies have been multiplied by $\beta = (k_BT)^{-1}$. }
   \label{SI:fig:energyComparison} 
\end{figure}

\clearpage 

\section{Intermembrane repulsion pressure }
\label{SI:sec:intermembrane_repulsion}

\begin{figure}[h!]
   \centering
   \includegraphics[width=0.6\columnwidth]{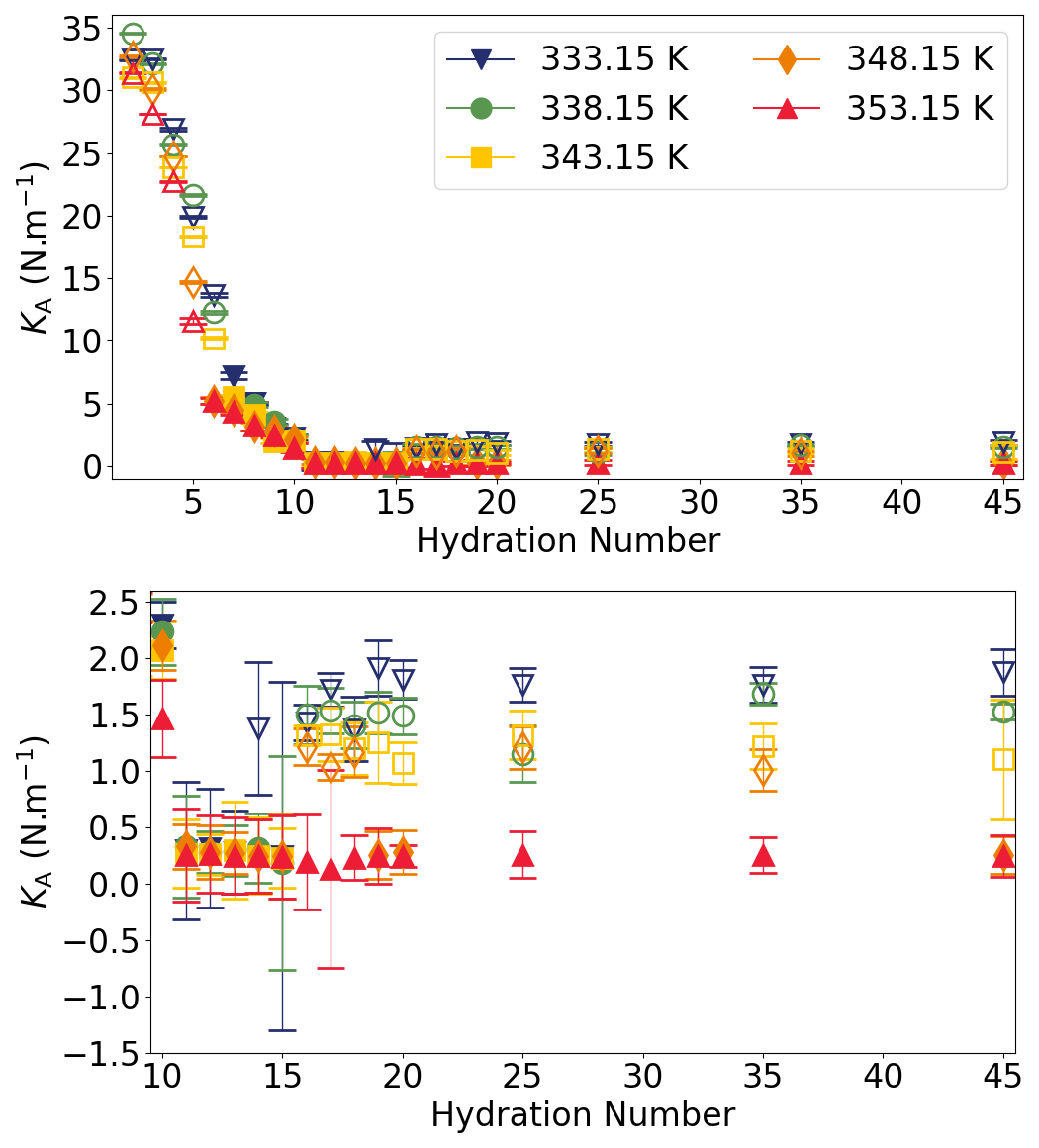}
   \caption{Area compressibility modulus as a function of HN at different temperatures, obtained from the fluctuations of area. Filled markers represent bilayers in the fluid phase and open markers represent bilayers in the gel phase. }
   \label{SI:fig:compressibilityModulus_vs_HN}
\end{figure}

\begin{figure}[h!]
   \centering
   \includegraphics[width=0.8\columnwidth]{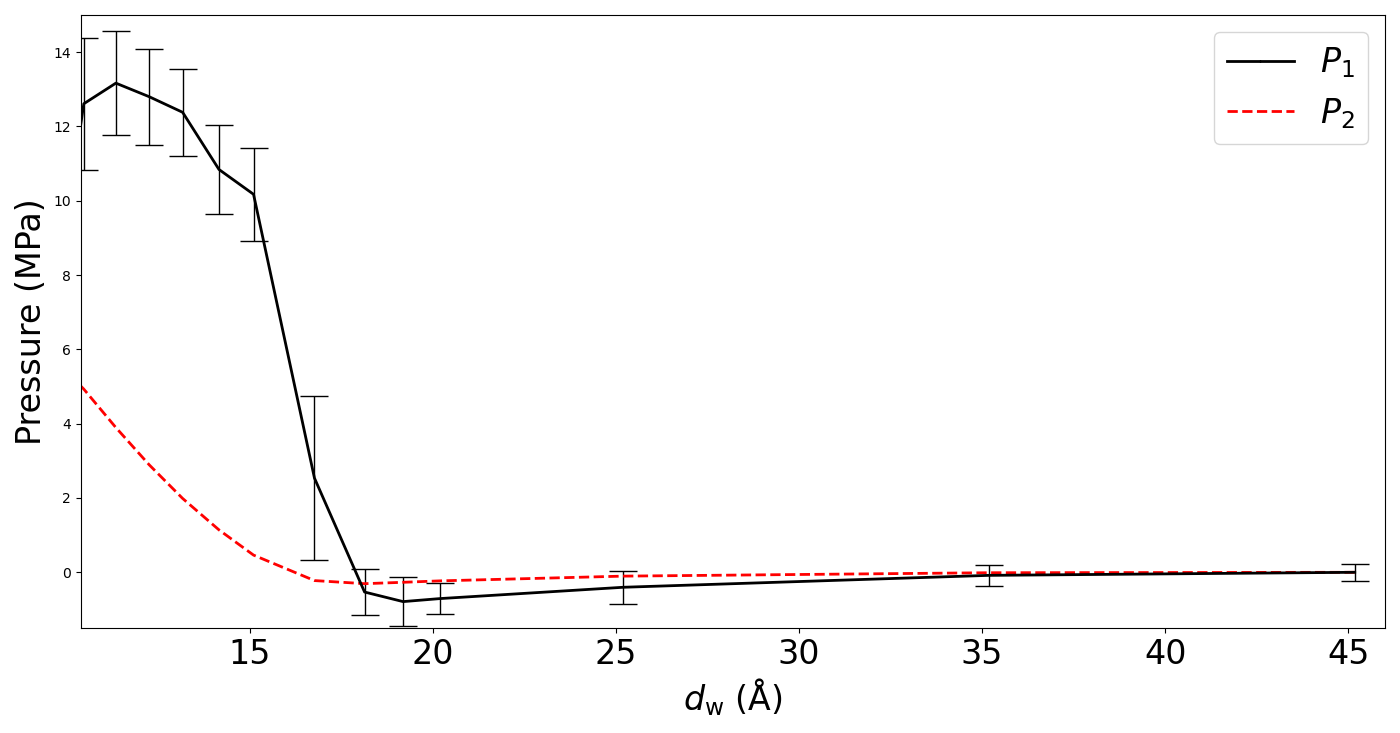}
   \caption{Values of the two pressure terms of Eq.~\eqref{eq:repulsionPressure}, as a function of water thickness  $d_{\mathrm{w}}$ obtained from Eq.~\eqref{eq:waterThicknessFromHN}, for the simulations at $T=353.15$~K : 
   $P_1(d_{{\mathrm w}}) = \frac{K_{\mathrm{A}}}{A_0} \frac{\Delta A(d_{\mathrm{w}})}{d_{\mathrm{w}}} $ 
   and 
   $P_2(d_{{\mathrm w}}) =
    \int_{d_{\mathrm{w}}}^{\infty} 
    \frac{P_1(d)}{d} 
    \mathrm{d}d
    $, so that $P(d_{{\mathrm w}}) = -
    \left[
    P_1(d_{{\mathrm w}})+P_2(d_{{\mathrm w}})
    \right]$. $P_2$ is negligible relative to $P_1$ for large hydration levels. For lower hydration ranges, $P_1$ remains of the same sign until \mbox{$10 \lq$~HN}, so that the signs of $P_1$  and $P_2$ are the same, so that one can use the sign of $P_1$ to infer the sign of $P$.}
   \label{SI:fig:P1_P2_vs_dw}
\end{figure}

\clearpage 

\section{Plots as a function of the water thickness} \label{SI:sec:plotsWaterThickness}

\begin{figure}[h!]
   \centering
   \includegraphics[width=0.6\columnwidth]{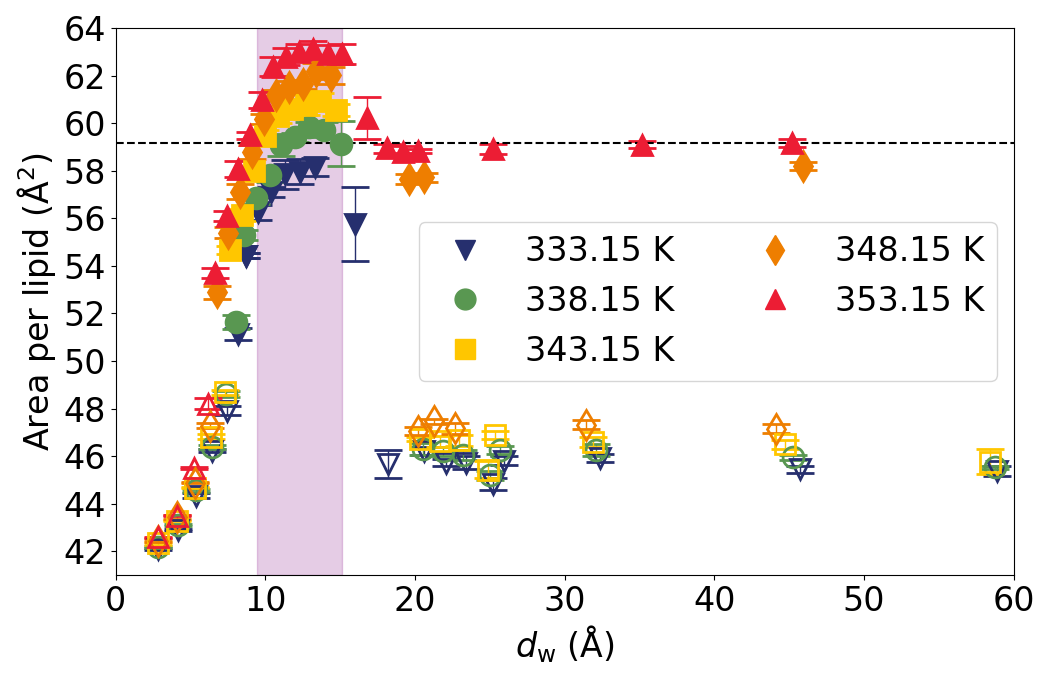}
   \caption{Area per lipid \textit{versus} water thickness $d_{\mathrm{w}}$ obtained from Eq.~\eqref{eq:waterThicknessFromHN}, at different temperatures. Filled markers represent bilayers in the fluid phase and open markers represent bilayers in the gel phase. The black dotted line shows the reference area per lipid $A_0$ at $T = 353.15$\,K. At this temperature, following Eq.~\eqref{eq:repulsionPressure}, we identify the purple region as the hydration range at which the attractive behaviour appears.}
   \label{SI:fig:ApL_vs_dw}
\end{figure}

\begin{figure}[h!]
   \centering
   \includegraphics[width=0.6\columnwidth]{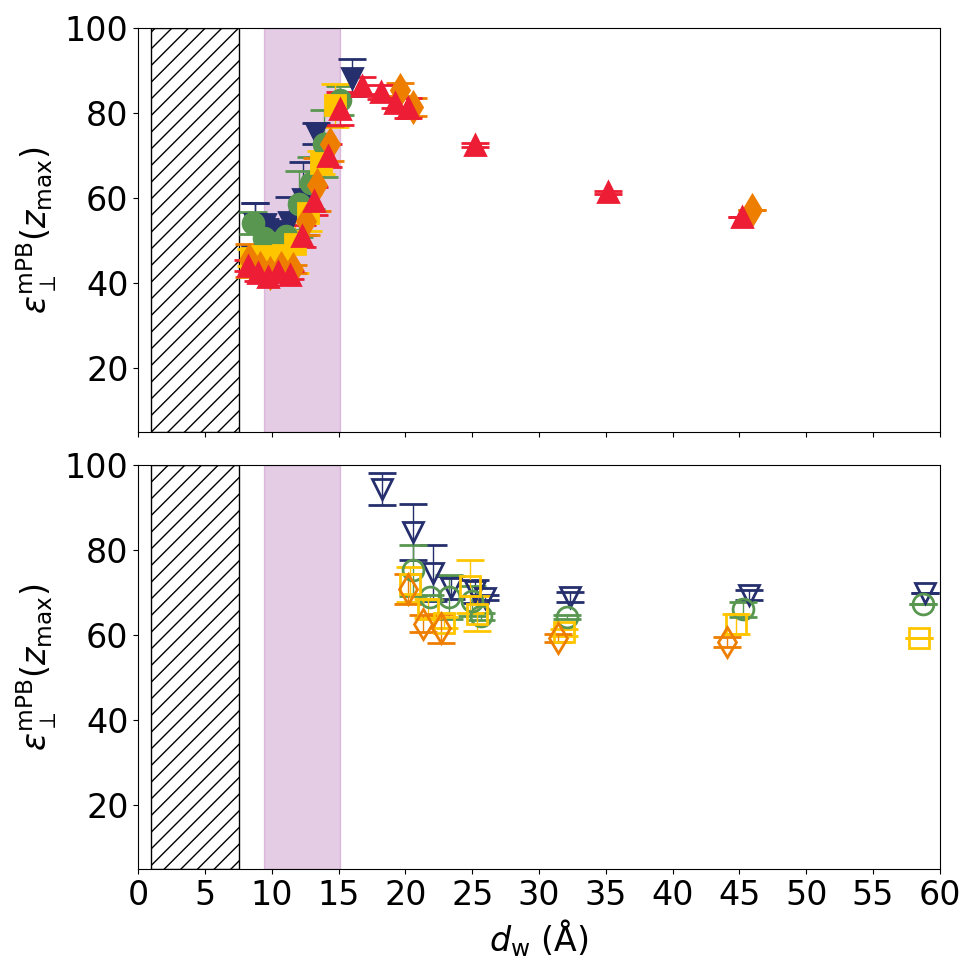}
   \caption{Maximum dielectric permittivity \textit{versus} water thickness $d_{\mathrm{w}}$ obtained from Eq.~\eqref{eq:waterThicknessFromHN}, at different temperatures, for bilayers in the fluid phase (top) and bilayers in the gel phase (bottom). The colours are the same as in Fig.~\ref{SI:fig:ApL_vs_dw} and the open/closed markers have the same meaning. The purple region represent where we identify an attractive regime between the membranes. The mPB model cannot be used in the hatched region because the ions no longer follow a Boltzmann distribution.} 
   \label{SI:fig:Epsilon_vs_dw}
\end{figure}

\begin{figure}[h!]
   \centering
   \includegraphics[width=0.6\columnwidth]{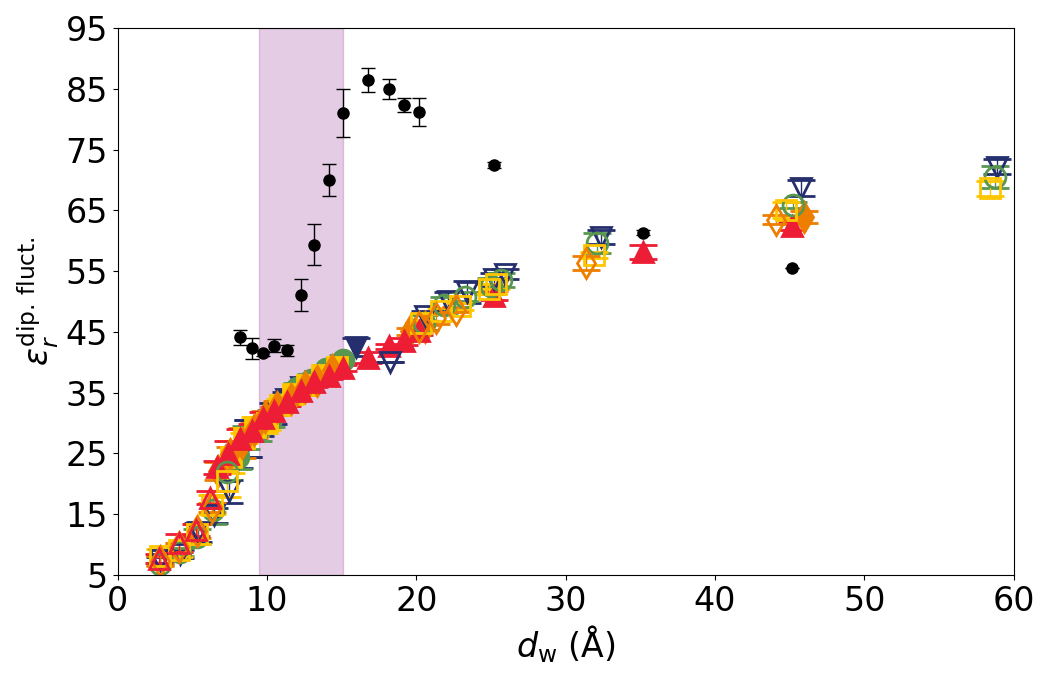}
   \caption{Dielectric response of the water molecules \textit{versus} water thickness $d_{\mathrm{w}}$ obtained from Eq.~\eqref{eq:waterThicknessFromHN}, at different temperatures. Black dots correspond to $\varepsilon_{\perp}^{\mathrm{mPB}}$ values at 353.15\,K. The colours are the same as in Fig.~\ref{SI:fig:ApL_vs_dw} and the open/closed markers have the same meaning. The purple region represent where we identify an attractive regime between the membranes.} 
   \label{SI:fig:EpsilonFromDipoles_vs_dw}
\end{figure}

\clearpage 

\section{Simulations List} 
\label{SI:sec:SimulationsList}

\begin{table}[h!]
    \centering
    \begin{tabular}{l  l l l l l l}
         HN &  DPPS & mTIP3P & Na$^{+}$ & Cl$^{-}$ & time (ns) & doi \\
         \hline
         45 &  200  & 9000   &  232  &  32  &   1000     &\texttt{10.5281/zenodo.16412906}\\
         35 &  200  & 7000   &  223  &  23  &   1000     &\texttt{10.5281/zenodo.16413739}\\
         25 &  200  & 5000   &  214  &  14  &   1000     & \texttt{10.5281/zenodo.16414106}\\
         20 &  200  & 4000   &  209  &  09  &   1000     &\texttt{10.5281/zenodo.16414722}\\
         19 &  200  & 3800   &  209  &  09  &   1000     & \texttt{10.5281/zenodo.16414851}\\
         18 &  200  & 3600   &  208  &  08  &   1000     & \texttt{soon available}\\
         17 &  200  & 3400   &  207  &  07  &   1000     & \texttt{soon available}\\
         16 &  200  & 3200   &  205  &  05  &   1000     & \texttt{soon available}\\
         15 &  200  & 3000   &  205  &  05  &   1000     & \texttt{10.5281/zenodo.16414919}\\
         14 &  200  & 2800   &  204  &  04  &   1000     & \texttt{soon available}\\
         13 &  200  & 2600   &  203  &  03  &   1000     & \texttt{soon available}\\
         12 &  200  & 2400   &  201  &  01  &   1000      & \texttt{soon available}\\
         11 &  200  & 2200   &  201  &  01  &   1000     & \texttt{soon available}\\
         10 &  800  & 8000   &  804  &  04  &    360     & \texttt{10.5281/zenodo.16318887}\\
         09 &  800  & 7200   &  804  &  04  &    360     & \texttt{soon available}\\
         08 &  800  & 6400   &  804  &  04  &    360     & \texttt{soon available}\\
         07 &  800  & 5600   &  803  &  03  &    360     & \texttt{soon available}\\
         06 &  800  & 4800   &  803  &  03  &    360     & \texttt{soon available}\\
         05 &  800  & 4000   &  802  &  02  &    360     & \texttt{soon available}\\
         04 &  800  & 3200   &  802  &  02  &    360     & \texttt{soon available}\\
         03 &  800  & 2400   &  801  &  01  &    360     & \texttt{soon available}\\
         02&   800  & 1600   &  801  &  01  &    360     & \texttt{soon available}\\
         & 
    \end{tabular}
    \caption{Composition of the systems for each HN, simulated at $T=$~333.15, 338.15, 343.15, 348.15 and 353.15\,K. The numbers correspond to the number of molecules in each system, or to the simulation time that was analyzed in the article. Lines with a doi have been published on Zenodo, with gromacs input files.  The simulations with the same composition are grouped into one doi, that contains the simulations at all the available temperatures.
    The trajectories used for the article have been analyzed every 10~ps. The ones on Zenodo are the same, but with configurations every 100~ps, to avoid very large files.}
    \label{SI:tab:SimulationsList}
\end{table}

%% file: 00_main.bbl
\begin{thebibliography}{10}

\bibitem{bib:Forbes1953}
R.~M. Forbes, A.~R. Cooper, and H.~H. Mitchell.
\newblock The composition of the adult human body as determined by chemical analysis.
\newblock {\em The Journal of biological chemistry}, 203:359--366, 1953.

\bibitem{bib:Binder2007}
H.~Binder.
\newblock Water near lipid membranes as seen by infrared spectroscopy.
\newblock {\em European Biophysics Journal}, 36:265--279, 2007.

\bibitem{bib:Levinger2002}
N.~E. Levinger.
\newblock Water in confinement.
\newblock {\em Science}, 298:1722--1723, 2002.

\bibitem{bib:Fumagalli2018}
L.~Fumagalli, A.~Esfandiar, R.~Fabregas, S.~Hu, P.~Ares, A.~Janardanan, Q.~Yang, B.~Radha, T.~Taniguchi, K.~Watanabe, G.~Gomila, K.~S. Novoselov, and A.~K. Geim.
\newblock Anomalously low dielectric constant of confined water.
\newblock {\em Science}, 360:1339--1342, 2018.

\bibitem{bib:BenTal1996}
N.~Ben-Tal, B.~Honig, R.~M. Peitzsch, G.~Denisov, and S.~McLaughlin.
\newblock Binding of small basic peptides to membranes containing acidic lipids: Theoretical models and experimental results.
\newblock {\em Biophysical Journal}, 71:561--575, 1996.

\bibitem{bib:Moreira2000}
A.~G. Moreira and R.~R. Netz.
\newblock Strong-coupling theory for counter-ion distributions.
\newblock {\em Europhysics Letters}, 52:705--711, 2000.

\bibitem{bib:Herrero2024}
C.~Herrero and L.~Joly.
\newblock The poisson-boltzmann equation in micro- and nanofluidics: A formulary.
\newblock {\em Physics of Fluids}, 36, 2024.

\bibitem{bib:Komorowski2018}
K.~Komorowski, A.~Salditt, Y.~Xu, H.~Yavuz, M.~Brennich, R.~Jahn, and T.~Salditt.
\newblock Vesicle adhesion and fusion studied by small-angle x-ray scattering.
\newblock {\em Biophysical Journal}, 114:1908--1920, 2018.

\bibitem{bib:Mukhina2019}
T.~Mukhina, A.~Hemmerle, V.~Rondelli, Y.~Gerelli, G.~Fragneto, J.~Daillant, and T.~Charitat.
\newblock Attractive interaction between fully charged lipid bilayers in a strongly confined geometry.
\newblock {\em Journal of Physical Chemistry Letters}, 10:7195--7199, 2019.

\bibitem{bib:Schlaich2019}
A.~Schlaich, A.~P.~Dos Santos, and R.~R. Netz.
\newblock Simulations of nanoseparated charged surfaces reveal charge-induced water reorientation and nonadditivity of hydration and mean-field electrostatic repulsion.
\newblock {\em Langmuir}, 35:551--560, 2019.

\bibitem{bib:Gramse2013}
G.~Gramse, A.~Dols-Perez, M.~A. Edwards, L.~Fumagalli, and G.~Gomila.
\newblock Nanoscale measurement of the dielectric constant of supported lipid bilayers in aqueous solutions with electrostatic force microscopy.
\newblock {\em Biophysical Journal}, 104:1257--1262, 2013.

\bibitem{bib:Schlaich2016}
A.~Schlaich, E.~W. Knapp, and R.~R. Netz.
\newblock Water dielectric effects in planar confinement.
\newblock {\em Physical Review Letters}, 117, 2016.

\bibitem{bib:Huang2008}
D.~M. Huang, C.~Cottin-Bizonne, C.~Ybert, and L.~Bocquet.
\newblock Aqueous electrolytes near hydrophobic surfaces: Dynamic effects of ion specificity and hydrodynamic slip.
\newblock {\em Langmuir}, 24:1442--1450, 2008.

\bibitem{bib:Poddar2016}
A.~Poddar, D.~Maity, A.~Bandopadhyay, and S.~Chakraborty.
\newblock Electrokinetics in polyelectrolyte grafted nanofluidic channels modulated by the ion partitioning effect.
\newblock {\em Soft Matter}, 12:5968--5978, 2016.

\bibitem{bib:Jo2008}
S.~Jo, T.~Kim, V.~G. Iyer, and W.~Im.
\newblock Charmm-gui: A web-based graphical user interface for charmm.
\newblock {\em Journal of Computational Chemistry}, 29:1859--1865, 2008.

\bibitem{bib:Klauda2010}
J.~B. Klauda, R.~M. Venable, J.~A. Freites, J.~W. O'Connor, D.~J. Tobias, C.~Mondragon-Ramirez, I.~Vorobyov, A.~D. MacKerell, and R.~W. Pastor.
\newblock Update of the charmm all-atom additive force field for lipids: Validation on six lipid types.
\newblock {\em Journal of Physical Chemistry B}, 114:7830--7843, 2010.

\bibitem{bib:MacKerell1998}
A.~D.~Jr. MacKerell, D.~Bashford, M.~Bellott, R.~L.~Jr. Dunbrack, J.~D. Evanseck, M.~J. Field, S.~Fischer, J.~Gao, H.~Guo, S.~Ha, D.~Joseph-McCarthy, L.~Kuchnir, K.~Kuczera, F.~T.~K. Lau, C.~Mattos, S.~Michnick, T.~Ngo, D.~T. Nguyen, B.~Prodhom, W.~E. Reiher, B.~Roux, M.~Schlenkrich, J.~C. Smith, R.~Stote, J.~Straub, M.~Watanabe, J.~Wiórkiewicz-Kuczera, D.~Yin, and M.~Karplus.
\newblock All-atom empirical potential for molecular modeling and dynamics studies of proteins.
\newblock {\em The Journal of Physical Chemistry B}, 102(18):3586--3616, 1998.

\bibitem{bib:Izadi2014}
S.~Izadi, R.~Anandakrishnan, and A.~V. Onufriev.
\newblock Building water models: A different approach.
\newblock {\em Journal of Physical Chemistry Letters}, 5:3863--3871, 2014.

\bibitem{bib:Vanommeslaeghe2015}
K.~Vanommeslaeghe and A.~D.~Mackerell Jr.
\newblock Charmm additive and polarizable force fields for biophysics and computer-aided drug design.
\newblock {\em Biochimica et Biophysica Acta - General Subjects}, 1850:861--871, 2015.

\bibitem{bib:DuboueDijon2020}
E.~Duboué-Dijon, M.~Javanainen, P.~Delcroix, P.~Jungwirth, and H.~Martinez-Seara.
\newblock A practical guide to biologically relevant molecular simulations with charge scaling for electronic polarization.
\newblock {\em Journal of Chemical Physics}, 153, 2020.

\bibitem{bib:Melcr2020}
J.~Melcr, T.~M. Ferreira, P.~Jungwirth, and O.~H.~S. Ollila.
\newblock Improved cation binding to lipid bilayers with negatively charged pops by effective inclusion of electronic polarization.
\newblock {\em Journal of Chemical Theory and Computation}, 16:738--748, 2020.

\bibitem{bib:Kurki2024}
Milla Kurki, Alexey~M. Nesterenko, Nicolai~E. Alsaker, Tiago M.~Ferreira, Sami Kyll{\"o}nen, Antti Poso, Piia Bartos, and Markus~S. Miettinen.
\newblock Solid-state nmr validation of opls4: Structure of pc-lipid bilayers and its modulation by dehydration.
\newblock {\em The Journal of Physical Chemistry B}, 128(50):12483--12492, 2024.

\bibitem{bib:Essman1995}
Ulrich Essmann, Lalith Perera, Max~L. Berkowitz, Tom Darden, Hsing Lee, and Lee~G. Pedersen.
\newblock A smooth particle mesh ewald method.
\newblock {\em The Journal of Chemical Physics}, 103(19):8577--8593, 11 1995.

\bibitem{bib:Darden1993}
Tom Darden, Darrin York, and Lee Pedersen.
\newblock Particle mesh ewald: An n.log(n) method for ewald sums in large systems.
\newblock {\em The Journal of Chemical Physics}, 98(12):10089--10092, 06 1993.

\bibitem{bib:Abraham2015}
M.~J. Abraham, T.~Murtola, R.~Schulz, S.~Páll, J.~C. Smith, B.~Hess, and E.~Lindahl.
\newblock Gromacs: High performance molecular simulations through multi-level parallelism from laptops to supercomputers.
\newblock {\em SoftwareX}, 1-2:19--25, 2015.

\bibitem{bib:Martyna1992}
Glenn~J. Martyna, Michael~L. Klein, and Mark Tuckerman.
\newblock Nosé–hoover chains: The canonical ensemble via continuous dynamics.
\newblock {\em The Journal of Chemical Physics}, 97(4):2635--2643, 08 1992.

\bibitem{bib:Parrinello1981}
M.~Parrinello and A.~Rahman.
\newblock Polymorphic transitions in single crystals: A new molecular dynamics method.
\newblock {\em Journal of Applied Physics}, 52(12):7182--7190, 12 1981.

\bibitem{bib:Israelachvili2011}
J.~N. Israelachvili.
\newblock {\em Intermolecular and Surface Forces (Third Edition)}.
\newblock Academic Press, 2011.

\bibitem{bib:Loche2019}
Philip Loche, Cihan Ayaz, Amanuel Wolde-Kidan, Alexander Schlaich, and Roland~R. Netz.
\newblock Universal and nonuniversal aspects of electrostatics in aqueous nanoconfinement.
\newblock {\em The Journal of Physical Chemistry B}, 124(21):4365--4371, 2020.
\newblock PMID: 32364728.

\bibitem{bib:Borgis2023}
D.~Borgis, D.~Laage, L.~Belloni, and G.~Jeanmairet.
\newblock Dielectric response of confined water films from a classical density functional theory perspective.
\newblock {\em Chemical Science}, 14:11141--11150, 2023.

\bibitem{bib:Stern2003}
H.~A. Stern and S.~E. Feller.
\newblock Calculation of the dielectric permittivity profile for a nonuniform system: Application to a lipid bilayer simulation.
\newblock {\em Journal of Chemical Physics}, 118:3401--3412, 2003.

\bibitem{bib:Nymeyer2008}
Hugh Nymeyer and Huan~Xiang Zhou.
\newblock A method to determine dielectric constants in nonhomogeneous systems: Application to biological membranes.
\newblock {\em Biophysical Journal}, 94:1185--1193, 2008.

\bibitem{bib:Huang1977}
Wei-Tze Huang and David~G. Levitt.
\newblock Theoretical calculation of the dielectric constant of a bilayer membrane.
\newblock {\em Biophysical Journal}, 17:111--128, 1977.

\bibitem{bib:Rashin1985}
Alexander~A. Rashin and Barry Honig.
\newblock Reevaluation of the born model of ion hydration.
\newblock {\em The Journal of Physical Chemistry}, 89:5588--5593, 1985.

\bibitem{bib:Kournopoulos2022}
Spiros Kournopoulos, Mirella~Simões Santos, Srikanth Ravipati, Andrew~J. Haslam, George Jackson, Ioannis~G. Economou, and Amparo Galindo.
\newblock The contribution of the ion-ion and ion-solvent interactions in a molecular thermodynamic treatment of electrolyte solutions.
\newblock {\em Journal of Physical Chemistry B}, 126:9821--9839, 2022.

\bibitem{bib:Silva2024}
Gabriel~M. Silva, Bjørn Maribo-Mogensen, Xiaodong Liang, and Georgios~M. Kontogeorgis.
\newblock Improving the born equation: Origin of the born radius and introducing dielectric saturation effects.
\newblock {\em Fluid Phase Equilibria}, 576, 2024.

\bibitem{bib:Schmid1999}
R.~Schmid, A.~M. Miah, and V.~N. Sapunov.
\newblock A new table of the thermodynamic quantities of ionic hydration: values and some applications (enthalpy–entropy compensation and born radii).
\newblock {\em Physical Chemistry Chemical Physics}, 2:97--102, 2000.

\bibitem{bib:Zhang2023}
Chunyi Zhang, Shuwen Yue, Athanassios~Z. Panagiotopoulos, Michael~L. Klein, and Xifan Wu.
\newblock Why dissolving salt in water decreases its dielectric permittivity.
\newblock {\em Phys. Rev. Lett.}, 131:076801, 08 2023.

\bibitem{bib:Gawrisch1992}
K.~Gawrisch, D.~Ruston, J.~Zimmerberg, V.~A. Parsegian, R.~P. Rand, and N.~Fuller.
\newblock Membrane dipole potentials, hydration forces, and the ordering of water at membrane surfaces.
\newblock {\em Biophysical Journal}, 61:1213--1223, 1992.

\bibitem{bib:Feller2000}
S.~E. Feller and A.~D. MacKerell.
\newblock An improved empirical potential energy function for molecular simulations of phospholipids.
\newblock {\em Journal of Physical Chemistry B}, 104:7510--7515, 2000.

\bibitem{bib:Malik2021}
S.~Malik and A.~Debnath.
\newblock Dehydration induced dynamical heterogeneity and ordering mechanism of lipid bilayers.
\newblock {\em Journal of Chemical Physics}, 154, 2021.

\bibitem{bib:Levin2002}
Y.~Levin.
\newblock Electrostatic correlations: from plasma to biology.
\newblock {\em Reports on Progress in Physics}, 65:1577--1632, 2002.

\bibitem{bib:Kirkwood1939}
J.~G. Kirkwood.
\newblock The dielectric polarization of polar liquids.
\newblock {\em The Journal of Chemical Physics}, 7:911--919, 1939.

\bibitem{bib:Miettinen2010}
Markus Miettinen.
\newblock {\em {Computational Modeling of Cationic Lipid Bilayers in Saline Solutions}}.
\newblock Aalto University, 2010.

\bibitem{bib:Gurtovenko2005}
Andrey~A. Gurtovenko, Markus Miettinen, Mikko Karttunen, and Ilpo Vattulainen.
\newblock Effect of monovalent salt on cationic lipid membranes as revealed by molecular dynamics simulations.
\newblock {\em The Journal of Physical Chemistry B}, 109(44):21126--21134, 2005.
\newblock PMID: 16853736.

\bibitem{bib:Miettinen2009}
Markus~S. Miettinen, Andrey~A. Gurtovenko, Ilpo Vattulainen, and Mikko Karttunen.
\newblock Ion dynamics in cationic lipid bilayer systems in saline solutions.
\newblock {\em The Journal of Physical Chemistry B}, 113(27):9226--9234, 2009.
\newblock PMID: 19534449.

\bibitem{bib:Marra1986}
J.~Marra.
\newblock Direct measurement of the interaction between phosphatidylglycerol bilayers in aqueous electrolyte solutions.
\newblock {\em Biophysical Journal}, 50:815--825, 1986.

\bibitem{bib:Schneck2012}
E.~Schneck, F.~Sedlmeier, and R.~R. Netz.
\newblock Hydration repulsion between biomembranes results from an interplay of dehydration and depolarization.
\newblock {\em Proceedings of the National Academy of Sciences of the United States of America}, 109:14405--14409, 2012.

\bibitem{bib:Schlaich2024}
A.~Schlaich, J.~O. Daldrop, B.~Kowalik, M.~Kanduč, E.~Schneck, and R.~R. Netz.
\newblock Water structuring induces nonuniversal hydration repulsion between polar surfaces: Quantitative comparison between molecular simulations, theory, and experiments.
\newblock {\em Langmuir}, 40:7896--7906, 2024.

\bibitem{bib:Smirnova2013}
Y.~G. Smirnova, S.~Aeffner, H.~J. Risselada, T.~Salditt, S.~J. Marrink, M.~Müller, and V.~Knecht.
\newblock Interbilayer repulsion forces between tension-free lipid bilayers from simulation.
\newblock {\em Soft Matter}, 9:10705--10718, 2013.

\bibitem{bib:Xu2010}
W.~Xu and F.~Pincet.
\newblock Quantification of phase transitions of lipid mixtures from bilayer to non-bilayer structures: Model, experimental validation and implication on membrane fusion.
\newblock {\em Chemistry and Physics of Lipids}, 163:280--285, 2010.

\bibitem{zenodo:gardre_2025_16280641}
Ludovic Gardré, 10.5281/zenodo.16318887 2025.

\bibitem{bib:Tieman2024}
Johanna~KS Tiemann, Magdalena Szczuka, Lisa Bouarroudj, Mohamed Oussaren, Steven Garcia, Rebecca~J Howard, Lucie Delemotte, Erik Lindahl, Marc Baaden, Kresten Lindorff-Larsen, Matthieu Chavent, and Pierre Poulain.
\newblock Mdverse, shedding light on the dark matter of molecular dynamics simulations.
\newblock {\em eLife}, 12:RP90061, 08 2024.

\bibitem{bib:Kiirikki2024}
Anne~M. Kiirikki, Hanne~S. Antila, Lara~S. Bort, Pavel Buslaev, Fernando Favela-Rosales, Tiago~Mendes Ferreira, Patrick~F.J. Fuchs, Rebeca Garcia-Fandino, Ivan Gushchin, Batuhan Kav, Norbert Kučerka, Patrik Kula, Milla Kurki, Alexander Kuzmin, Anusha Lalitha, Fabio Lolicato, Jesper~J. Madsen, Markus~S. Miettinen, Cedric Mingham, Luca Monticelli, Ricky Nencini, Alexey~M. Nesterenko, Thomas~J. Piggot, Ángel Piñeiro, Nathalie Reuter, Suman Samantray, Fabián Suárez-Lestón, Reza Talandashti, and O.~H.Samuli Ollila.
\newblock Overlay databank unlocks data-driven analyses of biomolecules for all.
\newblock {\em Nature Communications}, 15:1--15, 2024.

\end{thebibliography}
